\documentclass[preprint,12pt]{elsarticle}

\usepackage{fancyhdr}
\fancyheadoffset[RF]{0 pt}
\usepackage{graphicx}
\usepackage{graphics}
\usepackage{color}
\usepackage[dvipsnames]{xcolor}
\usepackage{array}
   \newcolumntype{P}[1]{>{\centering\arraybackslash}p{#1}}
\usepackage[center]{caption}
\usepackage{subcaption}
\usepackage{hyperref}
\usepackage{amsmath,bm}
\usepackage{bm}
\usepackage{amssymb}
\usepackage{amsfonts}
\usepackage[T1]{fontenc}
\usepackage[version=4]{mhchem}
\usepackage[noprefix]{nomencl}
\usepackage{xpatch}
\usepackage{siunitx}
\newcommand{\nomunit}[1]{%
\renewcommand{\nomentryend}{\hspace*{\fill}#1}}
\renewcommand\nomgroup[1]{%
  \item[\bfseries
  \ifstrequal{#1}{A}{Acronyms and subscripts}{%
  \ifstrequal{#1}{B}{Physical quantities}{%
  \ifstrequal{#1}{C}{Mathematical quantities}{}}}%
]}
\xpatchcmd{\thenomenclature}{%
  \section*{\nomname}
}{
}{\typeout{Success}}{\typeout{Failure}}
\makenomenclature
\RequirePackage{ifthen}
\renewcommand{\nomname}{\vspace{-0.5cm}}

\usepackage{tikz}
\usetikzlibrary{shapes}

\biboptions{numbers,sort&compress}

\usepackage{xcolor} 
\RequirePackage{fix-cm}

\journal{Combustion and Flame}

\begin{document}
%
\newcommand{\markerkonnovB}{\raisebox{0pt}{\tikz{\node[draw,scale=0.4,diamond,fill=black!10!black](){};}}}
\newcommand{\markerkonnovW}{\raisebox{0pt}{\tikz{\node[draw,scale=0.4,diamond,fill=white!10!white](){};}}}
\newcommand{\markerffcmB}{\raisebox{0pt}{\tikz{\node[draw,scale=0.3,regular polygon, regular polygon sides=3,fill=black!45!black](){};}}}
\newcommand{\markerffcmW}{\raisebox{0pt}{\tikz{\node[draw,scale=0.3,regular polygon, regular polygon sides=3,fill=white!45!white](){};}}}
\newcommand{\markersandiegoB}{\raisebox{0pt}{\tikz{\node[draw,scale=0.4,circle,fill=black!45!black](){};}}}
\newcommand{\markersandiegoW}{\raisebox{0pt}{\tikz{\node[draw,scale=0.4,circle,fill=white!45!white](){};}}}
\begin{frontmatter}

\title{From graph theory and geometric probabilities to a representative width for three-dimensional detonation cells}


\author{Vianney Monnier}
\author{Pierre Vidal*}
\author{Vincent Rodriguez}
\author{Ratiba Zitoun}

\address{Institut Pprime, UPR 3346 CNRS, Fluid, Thermal and Combustion Sciences Department,\\ENSMA, Téléport 2, 1 Av. Clément Ader, Chasseneuil-du-Poitou, 86360, France}
        
\cortext[ ]{Corresponding author: pierre.vidal@ensma.fr}

\begin{abstract}
We present a model for predicting a representative width for the three-dimensional irregular patterns observed on the front views of cellular detonation fronts in reactive gases. Its physical premise is that the cellular combustion process produces the same burnt mass per unit of time as the average planar steady Zel'dovich-von Neuman-D\"oring (ZND) process. The transverse waves of irregular cells are described as a stochastic system subject to a stationary ergodic process, considering that the distributions of the patterns should have identical temporal and statistical average properties. Graph theory then defines an ideal cell whose grouping is equivalent to the actual 3D cellular front, geometric probabilities determine the mean burned fraction that parameterizes the model, and ZND calculations close the problem with the time-position relationship of a fluid element in the ZND reaction zone. The model is limited to detonation reaction zones whose only ignition mechanism is adiabatic shock compression, such as those of the mixtures with \ce{H2}, \ce{C3H8} or \ce{C2H4} as fuels considered in this work. The comparison of their measured and calculated widths shows an agreement better than or within the accepted experimental uncertainties, depending on the quality of the chemical kinetic scheme used for the ZND calculations. However, the comparison for \ce{CH4}:\ce{O2} mixtures shows high overestimates, indirectly confirming that the detonation reaction zones in these mixtures certainly include other ignition mechanisms contributing to the combustion process, such as turbulent diffusion. In these situations, the cell widths on longitudinal soot recordings have a very large dispersion, so a mean width may thus not be a relevant detonation characteristic length. The model is easily implementable as a post-process of ZND profiles and provides rapid estimates of the cell width, length and reaction time.
\end{abstract}

\end{frontmatter}

\newpage
\section{\label{sec:Introduction}Introduction}

First identified experimentally in the late 1950s \cite{Denisov1959}, the cellular structure of the detonation reaction zone in gases is viewed today as an example of nonlinear instability of combustion waves in compressible reactive fluids, \textit{e.g.}, \cite{Ng2012_Zhang,Clavin2012,Clavin2016}. It is now recognized that its physical representation can only be three-dimensional. Experiments on constant-velocity detonations in straight tubes are the usual method to characterize the basic phenomenology of this structuring, namely a grouping of Mach waves consisting of forward propagating convex shocks bounded by transversely propagating shocks, respective to the tube axis. Experimental and numerical analyses, \textit{e.g.}, \cite{Monnier2022a,Crane2023}, evidence that the front views of these detonation cells usually form irregular polygonal patterns, in particular, if their number on the front surface is sufficiently large. That is observed typically in tubes with cross sections large enough because the usual cell descriptor, namely a mean width $\bar{\lambda}_\text{C}$, decreases as the initial pressure $p_0$ of the gas increases. The usual modelling framework is the hydrodynamics of an inviscid fluid undergoing adiabatic chemical evolution, although including viscosity, for example, from turbulent diffusion, appears necessary depending on the chemical composition, \cite{Soloukhin1969,Subbotin1975,Radulescu2005,Meagher2023}. A point of debate is also whether a single characteristic length is sufficient to represent these 3D irregular patterns. In the following, we assume that such is $\bar{\lambda}_\text{C}$, and, as a continuation of the analysis outlined from our experimental recordings of detonation cells \cite{Monnier2022a,Monnier2022b}, we propose a heuristic model for predicting it in the case of inviscid flows reacting adiabatically.

Experimental investigations are based mainly on recordings on soot-coated foils placed longitudinally against the inner wall(s) of the tube. The erosion of the coating by the interaction of the transverse and longitudinal shocks with the foil draws diamonds whose statistical analysis defines the mean width $\bar{\lambda}_\text{C}$. Mixtures of light fuels and oxygen highly diluted with a mono-atomic inert gas, such as argon, produce regular diamonds, and mixtures of heavy fuels and oxygen diluted with nitrogen produce more irregular ones \cite{StrehlowEngel1969,Libouton1981,Sharpe2008,Desbordes2012_Zhang}.
Thus, the diamonds are classified as very regular, regular, moderately regular, irregular and highly irregular, although this classification is considered to be subjective, especially for the less regular diamonds \cite{Sharpe2008,Zhao2016}. 
A non-dimensional indicator useful for quantifying the longitudinal regularity is the $\chi$ parameter whose values, \textit{e.g.}, \hyperref[sec:Results]{Sect. 5}, are calculated with the Zel'dovich-von Neuman-D\"oring (ZND) model of adiabatic planar detonation \cite{Higgins2012_Zhang}. Thus, the higher $\chi$, the more irregular the diamonds \cite{Short2003,Radulescu2003,Ng2012_Zhang,Radulescu2013}.
However, front-view recordings can be easily obtained from the impact of the detonation on a soot-coated foil placed perpendicular to the tube axis \cite{Denisov1959}. They show polygonal patterns, usually irregular, whose edges represent the positions of the transverse waves at the instant of impact. The observations indicate that these geometric properties depend markedly on the initial pressure $p_0$, temperature $T_0$ and mixture composition \cite{Stamps1991,Auffret1999,Auffret2001}, and the cross-section shape of the tube. Therefore, the geometrical properties of the diamonds alone are not sufficient to characterize the cellular structure, and the separation of the information into longitudinal and front-view recordings makes the combined analyses of the geometric properties of the diamonds and the front-view patterns difficult.

For example \cite{Monnier2022a}, for the mixture \ce{2H2 + O2 + 2Ar}, the front views show irregular polygons unless if the initial pressure is low enough and the tube cross section is square, in which case the front views show regular patterns of rectangles. The longitudinal recordings show rows of regular of diamonds regardless of $p_0$ and the cross-section shape. As $p_0$ decreases, the regular arrangement on the longitudinal recordings remains qualitatively the same in the round cross section, whereas lines emerge between the rows in the square cross section, representing the impact of transverse waves propagating perpendicular to the recording wall, coherent with the information from the front views.

The cell widths $\bar{\lambda}_\text{C}$ decrease with increasing $p_0$ and become independent of the cross-section shape if the number of cells per surface area is large enough, \textit{i.e.} when $p_0$ is greater than a limit that may depend on the cross-section area. Below this limit, the values of $\bar{\lambda}_\text{C}$ are smaller in the square cross section at the same $p_0$, with a difference that increases with decreasing $p_0$. The cross-section shape has no apparent effect on the cell aspect ratio above the limiting $p_0$, but the cells are slimmer in the square cross section below it. Thus, the cell dynamics at the walls of a tube at large enough $p_0$ or cell number may not represent that on the entire detonation front, and the information from longitudinal recordings alone may not be sufficient to describe the geometric properties of the cellular structure.
 
Marginal detonation regimes are observed for tube transverse dimensions or $p_0$ sufficiently small so that the number of cells on the front is not too large. Therefore, regular front-view patterns are exclusive of marginal detonations in mixtures with light fuels contained in tubes with square or rectangular cross sections. Irregular front-view patterns do not necessarily indicate multi-cellular regimes, \textit{e.g.}, Chapman-Jouguet (CJ), since this irregularity depends on the cross-section shape at sufficiently small $p_0$ and is inherent to to a sufficiently large number of cells.

The mean cell width $\bar{\lambda}_\text{C}$, determined from longitudinal recordings in straight tubes of sufficient length and width, has long been used as an indicator of detonability, defined as the ability of a detonation to propagate in systems with finite transverse dimensions. The smaller the cell width, the greater the detonability. The question is how relevant this criterion is for geometrically complex devices and how accurate this mean width can and should be. Due to experimental inaccuracies, the usual uncertainties on detonation cell widths are between 10\% and 25\%, \cite{Manzhalei1974,Bull1982,Moen1984,Tieszen1986,Aminallah1993}. We have also shown that no model can give an accuracy better than $\pm$ 14 \% \cite{Monnier2022b}. For highly irregular diamonds, the standard deviation is too high to make a mean width a representative length. Nevertheless, regardless of regularity, the mean width $\bar{\lambda}_\text{C}$ is often considered a characteristic length useful in detonation dynamics, for example, in the design of safe industrial processes and high-energy detonation devices. 

For mixtures whose longitudinal soot recordings show diamonds not too highly irregular, such as those with the fuels \ce{H2}, \ce{C2H4} or \ce{C3H8} mixed with \ce{O2}, the high-speed recordings indicate that the local combustion rate is much faster in the regions behind the transverse waves and the faster forward waves, that is, much faster than the mean cellular rate, \textit{e.g.}, \cite{Pintgen2003,Pintgen2003b,Austin2005,Frederick2022}. The dominant ignition mechanism is then adiabatic shock compression. In contrast, following earlier observations by Solhoukin \cite{Soloukhin1969} and Subbotin \cite{Subbotin1975}, recent analyses \cite{Short2003,Radulescu2003,Radulescu2013} point out that the cellular combustion process for highly irregular diamonds cannot be strictly adiabatic. The recordings show the participation of a distributed combustion mechanism from unburnt gas pockets much further away from the shock than the shock-induced adiabatic combustion zone. Two cellular combustion mechanisms are thus distinguished. The adiabatic mechanism gives regular to irregular diamonds, and its modelling is compatible with the Euler equations for reactive inviscid compressible fluids. The non-adiabatic mechanism gives highly irregular diamonds, and its modelling requires the Navier-Stokes equations, including turbulent diffusion, compressibility, and chemical kinetics.

Our model for predicting $\bar{\lambda}_\text{C}$ is for an adiabatic cellular process of a constant-velocity detonation propagating in a wide tube of constant cross-section so that the number of cells per unit area is large enough. As we have just discussed, this ensures the irregularity of the front-view patterns, but limits its predictive capacity to regular to irregular diamonds, \textit{i.e.} excludes the cases of highly-irregular diamonds. Its three ingredients are graph theory, geometric probabilities and the ZND model. Basic conservation law expresses the global equivalence of the adiabatic ZND and cellular combustion processes relative to their respective reaction times since the cellular process achieves the CJ velocity in a large and long detonation tube. The 3D cell front is considered to be a stochastic system in the stationary ergodic regime. Indeed, there is no physical reason why, sufficiently far from the ignition, the distributions of the geometric properties of irregular front-view patterns should be statistically different at one front position or another during the same experiment or, for the same initial conditions, at the same front position in one experiment or another.

In probability theory, stochasticity is non-determinism. That is the case with irregular front-view patterns due to their high sensitivity to initial and boundary conditions. No repetition of an experiment - physical or numerical - can achieve the same position, size or orientation of the same pattern, \textit{i.e.} a rectangle or a hexagon, at the same distance of the front from the ignition. Ergodicity is the hypothesis that sufficiently many elements governed by a stochastic process have identical temporal and statistical average properties, so a single realization of enough elements presumably gives the statistic. Correct predictions are usually the only support for this hypothesis. Here it means that the average properties of the patterns on the front at a given time are statistically the same as those obtained on all fronts up to that time. Ergodicity is stationary if the averages are independent of the sampling instant. That is the case for those of the irregular front-view patterns recorded far enough away from ignition in a tube sufficiently large. The samples are the positions, sizes and orientations of the patterns and hence the lengths and orientations of their edges. Thus, any front view farther in this experiment, or at the same position in other experiments, will achieve identical distributions, \textit{i.e.} statistically the same numbers of triangles, rectangles, hexagons, edges, etc.

The model is not intended to explain how detonation reaction zones become unstable, but rather to readily predict a physical cell mean width sufficiently accurate for practical purposes without detailing the complex wave interactions of the cellular structure.

In \hyperref[sec:mod-overview]{Section \ref{sec:mod-overview}}, we propose a summary of recent approaches to the characterization of the detonation cell width.
In \hyperref[sec:Model]{Section \ref{sec:Model}}, we express the physical premise that the unsteady 3D process for irregular front-view patterns should produce on average the same burnt mass as the steady planar ZND process per unit of time.
In \hyperref[sec:GraphProba]{Section \ref{sec:GraphProba}}, based on the ergodicity of the transverse-wave motion, we implement graph theory to define an ideal cell whose grouping is equivalent to the real 3D cellular front \cite{Monnier2022a} and geometric probabilities to determine the mean burnt mass fraction that parameterizes the model, and we close the problem of determining $\bar{\lambda}_{\text{C}}$ with the relation time-position of a fluid element in the ZND steady reaction zone.
In \hyperref[sec:Results]{Section 5} we implement the ZND model with several detailed chemical kinetic mechanisms to calculate mean cell widths for several mixtures depending on their initial pressure $p_0$ and equivalence ratio (ER), and we compare them to the measured values $\bar{\lambda}_{\text{C}}$.

In \hyperref[sec:DiscConcl]{Section \ref{sec:DiscConcl}}, we discuss the model, and we conclude.
In \hyperref[sec:App-ZND]{Appendix A}, we describe the equations of the adiabatic ZND model and their numerical integration, and, in \hyperref[sec:App-Nomen]{Appendix B}, we collect the notations. 

\clearpage 
\section{\label{sec:mod-overview}Cell width prediction: former approaches and ours}

Vasil'ev \cite{Vassiliev2012_Zhang} has proposed a classification into four groups of the approaches to cell width predictions. We suggest four different groups to reflect recent advances and situate our model.

The first group consists of high-resolution numerical simulations. Most are based on the Euler equations for reactive inviscid compressible fluids. They consider global or detailed chemical kinetic mechanisms but are often limited to two space dimensions. Such simplifications are helpful for quickly obtaining qualitative information on the sensitivity to the parameters of the reaction rates or the equation of state. However, parameter calibration for quantitative information is usually only possible on a narrow range of initial conditions, which limits their predictive ability for complex geometries. For example, numerical 2D cells may not have the correct width and aspect ratio. In the objective of design-cost reduction, restitution must not be confused with prediction. Nevertheless, the increase in computing capacity leads to a growing representativeness when considering the three dimensions of space and a detailed kinetic mechanism. For example, Crane et al. \cite{Crane2022} recently carried out simulations of front-view cellular patterns that, qualitatively, look strikingly like the experimental ones. That, however, still requires long computation times from the position of the ignition for a large cell number per unit area of the front surface. These drawbacks currently preclude the implementation of the necessary numerical tools, especially in the case of non-adiabatic subdomains or phenomena, such as transitions from deflagration to detonation, or very irregular cells, which requires solving the Navier-Stokes equations for a reactive compressible fluid (\hyperref[sec:Introduction]{Sect. 1}), \textit{e.g.}, \cite{Crane2019}. These approaches are likely to become the norm as soon as computing capacity allows because they can handle realistic equations of state and chemical kinetic mechanisms, just as an accurate calculation of CJ properties requires codes that take into account the dependence of the thermochemical properties of chemical species on pressure and temperature.

The second group comprises theoretical works that consider the ZND planar steady reaction zone as the initial condition from which small linear perturbations grow, \textit{e.g.}, \cite{JoulinVidal1998,Clavin2002,Clavin2012}. They aim to select the unstable modes that should evolve into the transverse shocks that represent the cellular front in the limits of long times and large transverse dimensions. They can combine the technique of matched asymptotic expansions and numerical analyses to obtain an evolution equation for the front surface. Its integration is expected to achieve a constant number of modes per unit area of the front surface, similar to multicellular detonations. Indeed, the experimental cells are independent of the shapes and the areas of the cross section in sufficiently long and wide detonation tubes \cite{Monnier2022a,Monnier2022b}. The selection of the salient parameters and magnitudes provides the main physics. However, they are still limited to two space dimensions. Also, a mathematical estimate of the long-time limit is difficult because the number of modes increases with increasing transverse dimensions, and too-small a computation box limits the number selected, similar to marginal detonations. 

The third group includes studies, partly numerical and theoretical, aimed at modelling the cellular front by mimicking the actual mechanism of transverse wave interactions. Early and recent analyses include those of Barthel \cite{Barthel1971,Barthel1974}, Strehlow \cite{Strehlow1967,Strehlowetal1969}, Crane \cite{Crane2021}, and Cheevers \cite{Cheevers2022}. The recent ones model the frontward-facing cell element as a cylindrical shock wave whose explosion kernel results from the collision of converging adjacent transverse waves. At some distance from the kernel, the reaction time becomes too large due to the combined effects of the transverse and longitudinal expansions. That defines the cell width as the chord of the diverging shock at that position. Their interest is to simplify the investigations in the second group. These approaches are ongoing and, to date, it is still difficult to assess what can be the cumulative effects of the assumptions on the shock dynamics and geometry, the kernel size and distributions and the constitutive relations.

\begin{figure}[h]
  \centering
  \includegraphics[width=0.5\linewidth]{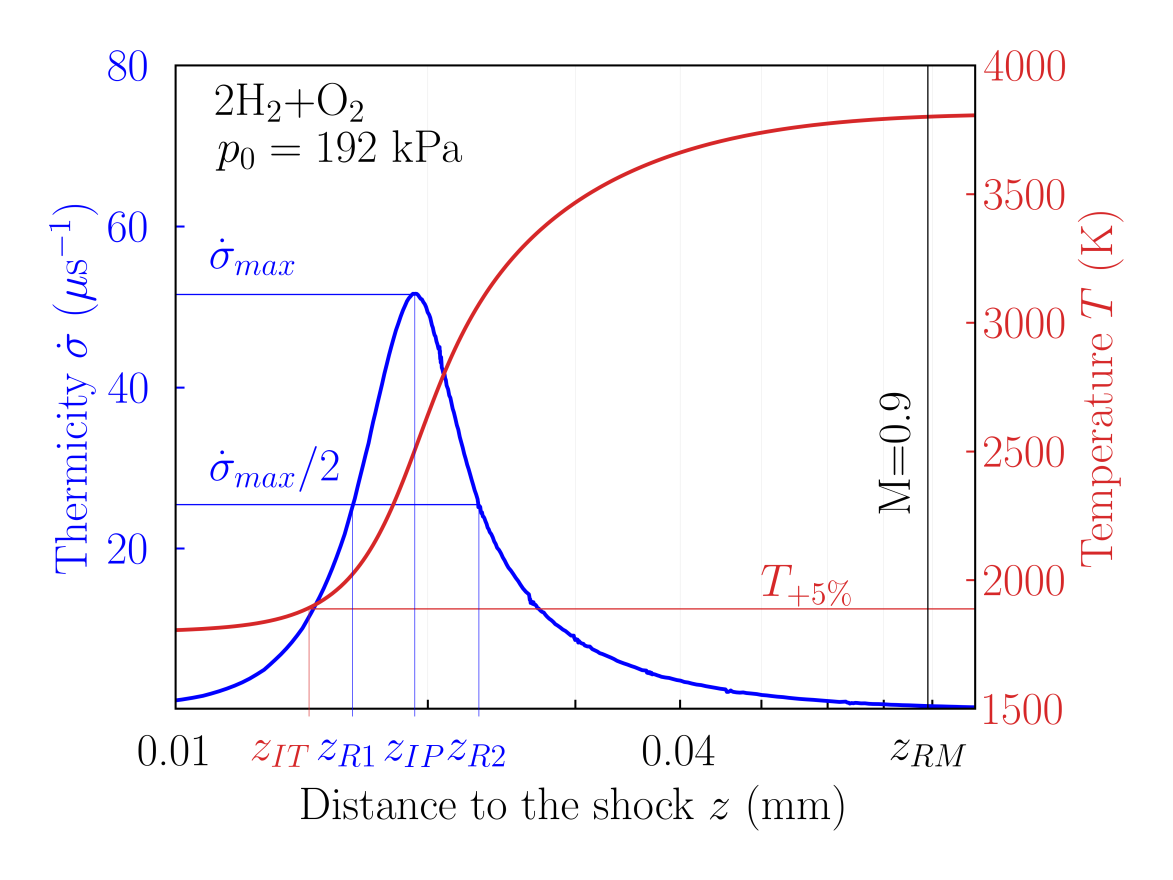}
  \caption{ZND profiles of thermicity (left scale) and temperature (right scale) illustrating several definitions of induction lengths, \textit{i.e.} $\bar{\ell}_\text{ZI}=z_\text{IT}$ or $z_\text{IP}$ and main-reaction thicknesses, \textit{i.e.} $\delta\bar{\ell}_\text{ZR}=z_\text{R2}-z_\text{R1}$ or $z_\text{RM}-z_\text{IT}$.}
  \label{fig:thermicite}
\end{figure}

The fourth group collects dimensional analyses based on a proportionality relationship between the cell mean width $\Bar{\lambda}_\text{C}$ and a characteristic length defined from the ZND profiles of temperature $T$ or thermicity $\dot{\sigma}$ (\hyperref[sec:App-ZND]{Apps. A} \hyperref[sec:App-Nomen]{\& B}). The original idea can be found in a work by Volin et al. \cite{Volin1960}. Depending on the authors, the characteristic length is the induction length $\bar{\ell}_\text{ZI}$ or a representative reaction length, $\bar{\ell}_\text{Z}$, which adds it with the thickness of main-reaction layer $\delta\bar{\ell}_\text{ZR}$, as suggested by, \textit{e.g.}, Shepherd \cite{Shepherd1986} and Gavrikov \cite{Gavrikov2000}, 
 \begin{align}
   &\Bar{\lambda}_\text{C}=k\;\bar{\ell}_\text{Z},\label{eq:K_lz}\\
   &\bar{\ell}_\text{Z}=\bar{\ell}_\text{ZI} \text{ or } \bar{\ell}_\text{Z}=\bar{\ell}_\text{ZI}+\delta\bar{\ell}_\text{ZR}. \label{eq:Correl_Prop}
 \end{align}
The definitions of $\bar{\ell}_\text{ZI}$ and $\delta\bar{\ell}_\text{ZR}$ vary between authors, and Figure \ref{fig:thermicite} illustrates some.

Mathematically, the induction zone should be characterized from an asymptotic analysis, so the induction time is $1/(d\dot\sigma/dt)/\dot\sigma)_\text{N}$ and the induction length $U_\text{N}/(d\dot\sigma/dt)/\dot\sigma)_\text{N}$, where the subscript N denotes the state at the ZND shock. With the notations of the figure \ref{fig:thermicite}, practical estimates of $\bar{\ell}_\text{ZI}$ are the distances from the shock to the position where the temperature rise is $5\%$ \cite{Crane2019} ($z_\text{IT}$) or the thermicity is maximum \cite{Higgins2012_Zhang} ($z_\text{IP}$) - the position of the maximum temperature gradient is usually indistinguishable from $z_\text{IP}$. Short and Sharpe \cite{Short2003} define the main-reaction thickness $\delta\bar{\ell}_\text{ZR}$ as the distance between the positions where the Mach number reaches 0.9 and where the temperature gradient is maximum ($z_\text{RM}-z_\text{IP}$). Radulescu et al. \cite{Radulescu2003} and Ng and Zhang \cite{Ng2012_Zhang} define it by the ratio $U_\text{*}/\dot\sigma_\text{max}$, \textit{i.e.} the product of a characteristic material velocity and a characteristic time equal to the inverse of the maximum thermicity. Radulescu et al. take $U_\text{*}$ equal to the material velocity at the position of the thermicity maximum, while Ng and Zhang take that of the burnt gases in the CJ state. Crane et al. \cite{Crane2019} take $\delta\bar{\ell}_\text{ZR}$ equal to the width of the thermicity profile at its half-height ($z_\text{R2}-z_\text{R1}$). The choices of the induction and main-reaction thicknesses should exclude disjunctions and intersections. These choices are made a priori, so the predictive use of the proportionality relation (\ref{eq:Correl_Prop}) first necessitates fitting the coefficient $k$ to measured cell widths and building a table for categories of mixtures. The ratio $k=\Bar{\lambda}_\text{C}/\bar{\ell}_\text{Z}$ can therefore be quite sensitive to the choice of $\bar{\ell}_\text{Z}$ or the chemical kinetic mechanism. For example, for near stoichiometric mixtures whose fuel is \ce{H2}, \ce{C2H2} or \ce{C2H4} and the oxidant is oxygen and air, $k\approx29$ when $\bar{\ell}_\text{Z}=\bar{\ell}_\text{ZI}$, and $k\approx10$ when $\bar{\ell}_\text{Z}=\bar{\ell}_\text{ZI}+\delta\bar{\ell}_\text{ZR}$, \textit{e.g.}, \cite{Crane2019}. This approach certainly has the advantage of simplicity and physical sense, but relation (\ref{eq:Correl_Prop}) is a correlation whose predictive operability requires prior cell measurements. Its functional form, proportionality coefficient $k$, and characteristic length $\bar{\ell}_\text{Z}$ are not deduced jointly and explicitly from an analysis of the actual cellular mechanisms.

Our approach is a synthetic intermediate between the third and fourth groups. Its contribution consists in a closed system of equations that jointly determines the average geometric properties of the cell - its front-view pattern, the length, width and aspect ratio of the diamonds - and the associated ZND characteristic lengths - induction and main-reaction thicknesses ($\bar{\ell}_\text{ZI}$ and $\delta\bar{\ell}_\text{ZR}$) and representative reaction length ($\bar{\ell}_\text{Z}$) - without postulating one to obtain others and simplifying the chemical kinetic mechanism. Its basis is the opposite of that of the second group. The front is multicellular in the sense of both limits of long times and large transverse dimensions, namely a large set of irregular patterns that behave as a stationary ergodic system. That leads to the probability analysis in \hyperref[sec:GraphProba]{Section 4} without resorting to simplifying assumptions similar to those of the third group. Compared to the latter, our model establishes the proportionality relation (\ref{eq:K_lz}) based on the necessary global equivalence of the cellular and ZND reaction rates and determines the proportionality coefficient $k$ as a function of the shock and CJ properties and the value of the mean burnt mass fraction given by the probability analysis.

In essence, our model implements a probabilistic approach expressing that the cellular and ZND mechanisms should have equivalent combustion processes so that the cellular front can propagate at the same average velocity as the ZND front. Its validity is limited to multicellular detonations, not marginal ones, and an adiabatic cellular mechanism.

\clearpage
\section{\label{sec:Model}Model}

The basic assumption is that the cellular and ZND processes burn the fresh mixture at the same mass rate for sufficiently long periods. The ZND model of this work is that for constant-velocity detonation, \textit{i.e.} a steady planar reaction zone induced by a planar shock of constant velocity $D$. This ZND process is a valid average of the cellular process if the number of cells is very large. We choose as the origin of distances $x$ in the reference frame of the laboratory an arbitrary initial position $L_0=0$ of the ZND front and as the origin of times $t$ the instant $t_0=0$ when the fluid elements enter the ZND reaction zone. Thus, at an instant $t>0$, the position of the front is $L(t)=Dt$ and the distance measured from the front is $z=Dt-x$. Steadiness is the invariance of any variable $f$ with respect to $L(t)$ at a constant relative distance $z$ in the reaction zone, so $f$ is function of a single independent variable, namely $t$ or $z$. The relative position $z_\text{m}(t)$ of a fluid element, or the period $t_\text{m}(z)$ elapsed from the front to this position, are calculated by integrating the material speed with the ZND equations (App. A).

We refer to below as a complete reaction time the period necessary to completely burn all fluid elements captured by a front at $t_0=0$ and through the same reference surface area. Let $\bar{t}_{\text{C}}$ the period during which the ZND front travels the distance $\bar{L}_\text{C}$ representing the length of the mean cell. For the self-sustained detonation propagating at the Chapman-Jouguet (CJ) velocity $D_{\text{CJ}}$,
\begin{equation}
\bar{L}_\text{C}=D_{\text{CJ}}\,\bar{t}_{\text{C}}.  \label{Cell Length}
\end{equation}
\begin{figure*}[ht!]
\centering
\includegraphics[width=0.655\linewidth]{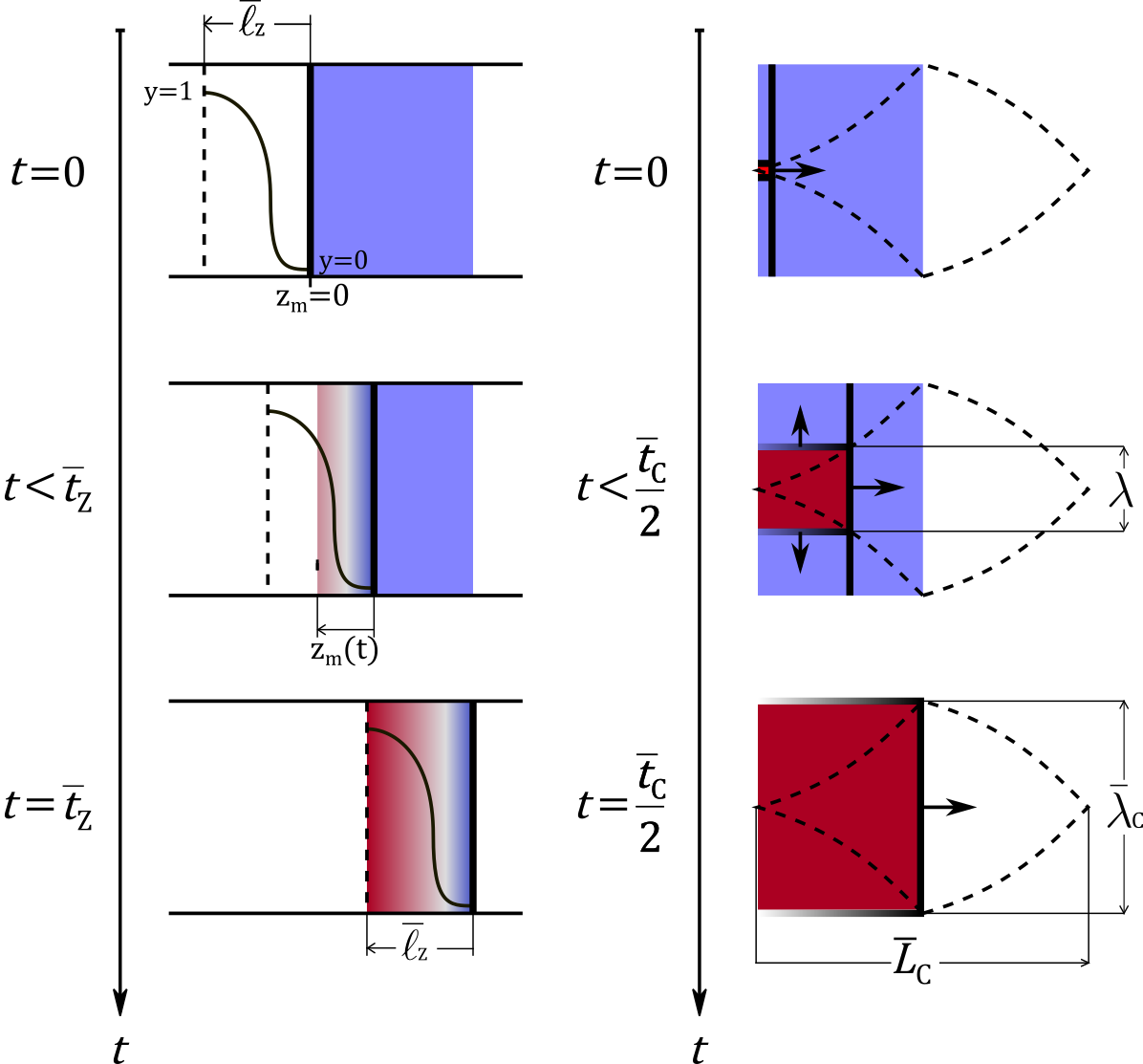}
\caption{Left: ZND process. Right: model adiabatic cellular process
\label{fig:ZND_Cell}
}
\vspace{-0.1cm} 
\end{figure*}
In the ZND process, denoting by $\bar{t}_{\text{Z}}$ its complete reaction time, the fluid elements entered in the reaction zone during the period $0<t\leqslant \bar{t}_{\text{Z}}$ can only be partially burnt at $\bar{t}_{\text{Z}}$ (Fig. \ref{fig:ZND_Cell}-left). That results in the mean ZND burnt mass fraction $\bar{y}_{\text{Z}}$ and reaction rate $\bar{y}_{\text{Z}}/\bar{t}_{\text{Z}}$. In the cellular process, the front is a grouping of forward-convex waves whose forefront velocities for irregular cells vary cyclically but randomly about the ZND mean velocity, such as $D_{\text{CJ}}$, \textit{e.g.}, \cite{VanTiggelen1989}. Their boundaries are the intersections with transverse waves that sweep the surfaces of the slower forward waves. For mixtures whose longitudinal soot recordings show diamonds with not too high irregularity, such as those with the light fuel \ce{H2} or the heavier hydrocarbon fuels \ce{C2H4} or \ce{C3H8}, the high-speed recordings indicate that the local combustion rate is much faster in the regions behind the transverse waves and the faster forward waves, that is, much faster than the mean cellular rate, \textit{e.g.}, \cite{Pintgen2003,Pintgen2003b,Austin2005,Frederick2022}. The dominant ignition mechanism is then the adiabatic shock compression, and a limiting symmetry argument then suggests that the complete reaction time, as defined above, of the ideal cell should be half the cell time $\bar{t}_{\text{C}}/2$. Indeed, the period $[0,\bar{t}_{\text{C}}/2]$ is that necessary, on average, for the transverse waves to sweep a projected front area equivalent to the maximum area of the ideal cell, which, by symmetry, occurs every cell half length $\bar{L}_{\text{C}}/2$. Thus, during this period, they cover the surface of the ideal cell, and they can capture and burn all the fluid elements that have crossed the lower-velocity front surfaces since $t_0$ (Fig. \ref{fig:ZND_Cell}-right). That results in the mean cell burnt mass fraction $\bar{y}_{\text{C}}$ and reaction rate $2/\bar{t}_{\text{C}}$ -- and not $\bar{y}_{\text{C}}\;2/\bar{t}_{\text{C}}$. The means of the mass fractions $y_{\text{Z}}$ and $y_{\text{C}}$ are relative to periods elapsed since $t_0=0$. They write
\begin{equation}
\bar{y}_{\text{Z}}=\frac{1}{\bar{t}_{\text{Z}}}\int_{0}^{\bar{t}_{\text{Z}}}y_{\text{Z}%
}(t)dt ,\quad \bar{y}_{\text{C}}=\frac{2}{\bar{t}_{\text{C}}}%
\int_{0}^{\bar{t}_{\text{C}}\hspace{-0.03cm}/\hspace{-0.02cm}2}y_{\text{C}%
}(t)dt ,  \label{Time Averages}
\end{equation}
where the subscripts Z and C denote the ZND and the cellular processes. The first time average above also applies to any variable in the ZND reaction zone, for example, the material speed ${U_{\text{Z}}}\left(t\right) =dz_\text{m}\left(t\right)/dt$ at the instant $t$ or the position $z_\text{m}\left(t\right)$ of a fluid element. This defines the ZND complete reaction length $\bar{\ell}_{\text{Z}}$ by
\begin{equation}
\bar{\ell}_{\text{Z}}=\int_{0}^{\bar{t}_{\text{Z}}}U_{\text{Z}}(t)dt =\bar{U}_{\text{Z}%
} \bar{t}_{\text{Z}},\quad \bar{U}_{\text{Z}}=\frac{\bar{\ell}_{\text{Z}}}{\bar{t}_{\text{Z}}},  \label{ZND Length}
\end{equation}
where $\bar{U}_{\text{Z}}$ denotes the mean of ${U_{\text{Z}}}\left(t\right) $. With $v$ denoting the specific volume, and $v_{0}$ its initial value, the relation of mass conservation for the ZND process (App. A)
\begin{equation}
{v_{\text{Z}}}D_{\text{CJ}}=v_{0}{U_{\text{Z}}}, \label{ZND Mass Conserv}
\end{equation}
valid at any position $z_\text{m}(t)$, can also be averaged, and (\ref{ZND Length}) and (\ref{ZND Mass Conserv}) rewrite
\begin{align}
\bar{\ell}_{\text{Z}} &=\frac{\bar{v}_{\text{Z}}}{v_{0}}D_{\text{CJ}}\; \bar{t}_{\text{Z}},  \label{ZND Length 2} \\
\bar{v}_{\text{Z}}\left( \bar{y}_{\text{Z}}\right) &=\left( 1-\bar{y}_{\text{Z}}\right) v_{\text{N}}+\bar{y}_{\text{Z}}v_{\text{CJ}},  \label{ZND Average Volume} \\
\bar{U}_{\text{Z}}\left( \bar{y}_{\text{Z}}\right)&=\left( 1-\bar{y}_{\text{Z}}\right) U_{\text{N}}+\bar{y}_{\text{Z}}U_{\text{CJ}}.  \label{ZND Average Velocity}
\end{align}
The relation (\ref{ZND Average Volume}) results from (\ref{ZND Mass Conserv}) and the averaging of the volume additivity relation $v=\sum y_{i}v_{i}$, where $v_{i}$ and $y_{i}$ denote the specific volume and the mass fraction of the chemical species $i$, and $v_{\text{N}}$ and $v_{\text{CJ}}$ denote the specific volumes at the ZND shock (subscript N) and the reaction end positions (subscript CJ for a self-sustained reaction zone, \textit{i.e.} this work). The ZND reaction zone thus reduces to an induction layer without chemical reactions and a main reaction layer that concentrates the burnt mass. Per unit of surface area, these masses are $M_{\text{N}}=(U_{\text{N}}/v_{\text{N}})\bar{t}_{\text{ZI}}$ and $M_{\text{B}}=(U_{\text{CJ}}/v_{\text{CJ}})\delta \bar{t}_{\text{ZR}}$ respectively, where $\bar{t}_{\text{ZI}}$ and $\delta \bar{t}_{\text{ZR}}$ are the crossing times of these layers ($\bar{t}_{\text{Z}}=\bar{t}_{\text{ZI}}+\delta \bar{t}_{\text{ZR}}$). The mean fraction $\bar{y}_{\text{Z}}$ is the ratio $M_{\text{B}}/M_{\text{Z}}$, where $M_{\text{Z}}=M_{\text{N}}+M_{\text{B}} =(\bar{U}_{\text{Z}}/\bar{v}_{\text{Z}})\bar{t}_{\text{Z}}$ is the total mass in the ZND reaction zone, and it is also the ratio $\delta \bar{t}_{\text{ZR}}/\bar{t}_{\text{Z}}$ since $\bar{U}_{\text{Z}}/\bar{v}_{\text{Z}}$ is constant (\ref{ZND Mass Conserv}).

\begin{figure*}[ht!]
\centering
  \begin{subfigure}{.5\textwidth}
  \centering
  \includegraphics[width=1\linewidth]{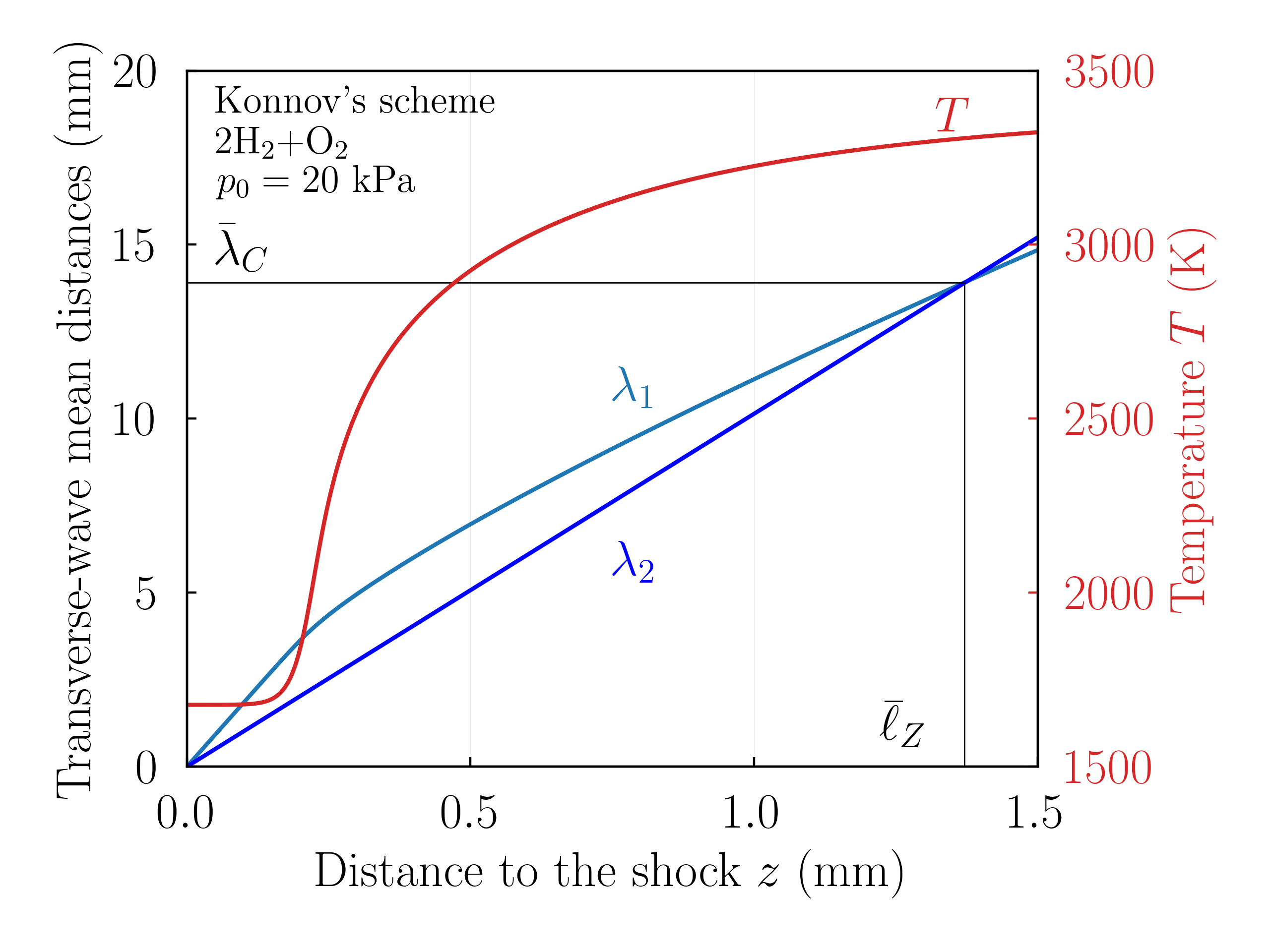}
  \end{subfigure}
\hspace{-0.25cm}
  \begin{subfigure}{.5\textwidth}
  \centering
  \includegraphics[width=1\linewidth]{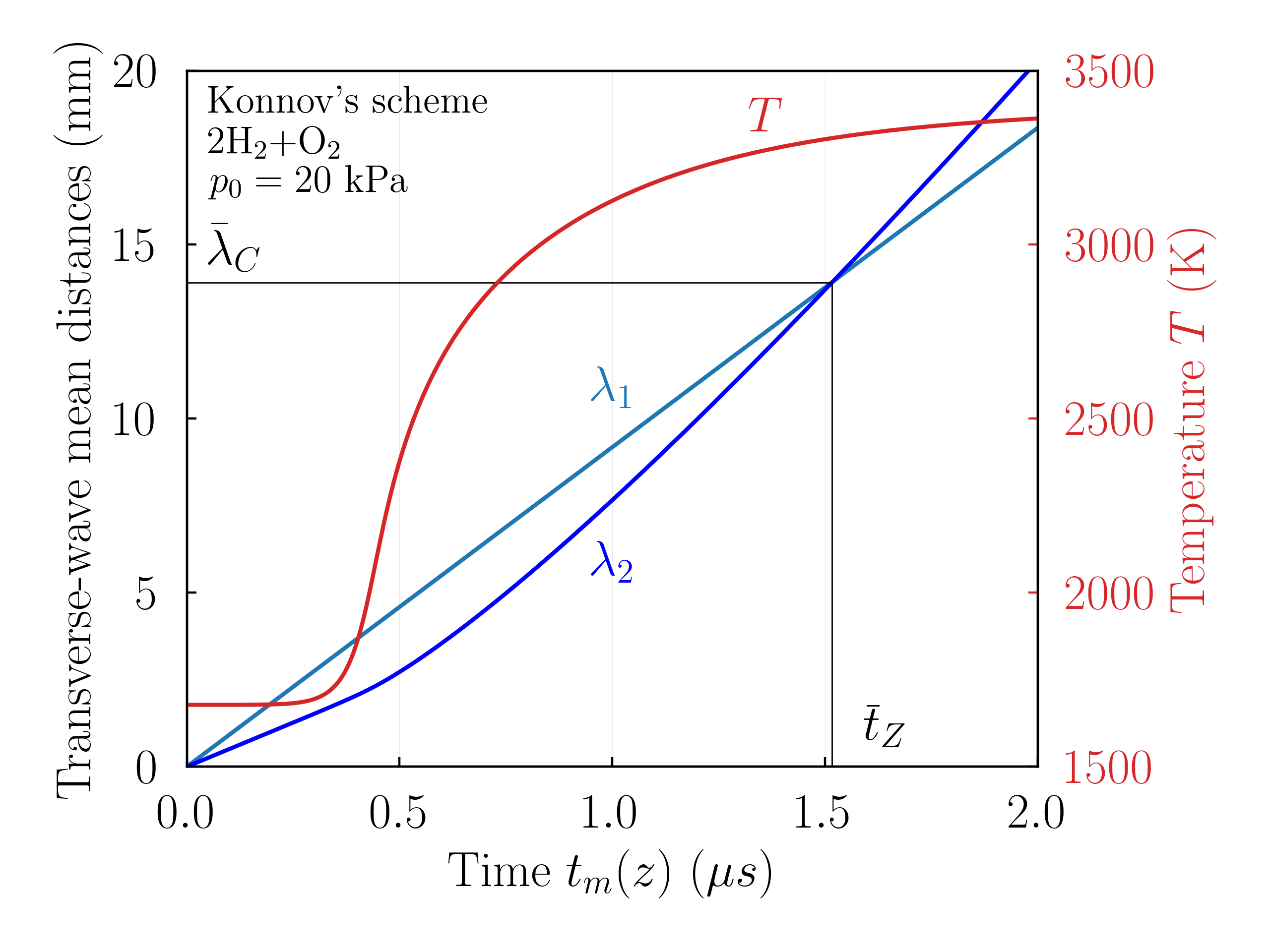}
  \end{subfigure}
\caption{ZND profile (left graph) and material evolution (right graph) of the temperature $T$ (red curves, right ordinates) and the transverse-wave mean distances ${\lambda}_1$ and ${\lambda}_2$ (blue and green curves, left ordinates). The intersections of ${\lambda}_1$ and ${\lambda}_2$ give the cell mean width $\bar{\lambda}_\text{C}$ and the ZND reaction length $\bar{\ell}_\text{Z}$ and time $\bar{t}_\text{Z}$}.
\label{fig:ZND_Intersections}
\end{figure*}

The equality of the cellular and ZND average reaction rates implies that of their average reaction progress variables $\bar{y}_{\text{C}}$ and $\bar{y}_{\text{Z}}$ with respect to their reaction times $\bar{t}_{\text{C}}/2$ and $\bar{t}_{\text{Z}}$, so
\begin{equation}
\frac{2}{\bar{t}_{\text{C}}} =\frac{\bar{y}}{\bar{t}_{\text{Z}}}, \label{Eq Rate}
\end{equation}
where $\bar{y}$ denotes $\bar{y}_{\text{C}}=\bar{y}_{\text{Z}}$. The combination of (\ref{Cell Length}) with (\ref{Eq Rate}) gives the relation (\ref{Cell Length 1}) between the cell mean length $\bar{L}_\text{C}$ and the ZND complete reaction time $\bar{t}_{\text{Z}}$, which, with (\ref{ZND Length 2}) and (\ref{ZND Average Volume}), gives
the relation (\ref{Cell Length 2}) between $\bar{L}_{\text{C}}$ and the ZND complete reaction length $\bar{\ell}_{\text{Z}}$,
\begin{align}
\hspace{3.5cm}\bar{L}_\text{C}\left( \bar{y},\bar{t}_{\text{Z}}\right) &=k_{1}\; \bar{t}_{\text{Z}%
},  &k_{1}\left( \bar{y}\right) &=\frac{2}{\bar{y}}D_{\text{CJ}},
\hspace{3.5cm} \label{Cell Length 1} \\
\hspace{3.5cm}\bar{L}_\text{C}\left( \bar{y},\bar{\ell}_{\text{Z}}\right) &=k_{2}\; \bar{\ell}_{%
\text{Z}}, &k_{2}\left( \bar{y}\right) &=\frac{2}{\bar{y}}\frac{v_{0}}{%
\bar{v}_{\text{Z}}\left( \bar{y}\right) }.  \hspace{3.5cm} \label{Cell Length 2}
\end{align}
In the next section (\hyperref[sec:GraphProba]{Sect. 4}), graph theory is used to define a cellular pattern statistically equivalent to those on an irregular 3D cellular front. Geometric probabilities applied to this ideal cell give the value of the mean burnt mass fraction $\bar{y}\approx 0.385$, and geometry considerations give the value of the width-to-length aspect ratio $a=\bar{\lambda}_\text{C}/\bar{L}_\text{C}\approx 0.64$ (\hyperref[sec:GraphProba]{Sect. 4}).
Then the cell mean width $\bar{\lambda}_\text{C}$ and length $\bar{L}_\text{C}$, and the ZND characteristic length $\bar{\ell}_{\text{Z}}$ and time $\bar{t}_{\text{Z}}$ are determined by the intersection of the curves $\lambda_1\left(z\right)$ and $\lambda_2\left(z\right)$,
\begin{align}
\lambda_1\left(z\right) &= a\; k_{1}\; t_{\text{m}}(z), \label{Cell Width 1 z}\\
\lambda_2\left(z\right) &= a\; k_{2}\; z, \label{Cell Width 2 z}
\end{align}
or the curves  $\lambda_1\left(t\right)$ and $\lambda_2\left(t\right)$,
\begin{align}
\lambda_1\left(t\right) &= a\; k_{1}\; t, \label{Cell Width 1 t}\\
\lambda_2\left(t\right) &= a\; k_{2}\; z_{\text{m}}(t), \label{Cell Width 2 t}
\end{align}
which Figure \ref{fig:ZND_Intersections} illustrates. The intersection of $\lambda_1$ and $\lambda_2$ expresses the compatibility constraint that the cellular process will generate the same mean burnt mass fraction at $\Bar{t}_\text{C}$ as the ZND process at $\bar{t}_{\text{Z}}=t_\text{m}(z=\bar{\ell}_{\text{Z}})$ or, equivalently, at $\bar{\ell}_{\text{Z}}=z_\text{m}(t=\bar{t}_{\text{Z}})$. The function $t_{\text{m}}(z)$ in (\ref{Cell Width 1 z}) is the relation time-position of a fluid element in the ZND reaction zone, \textit{i.e.} the time it takes a fluid element to reach the position $z$. The function $z_{\text{m}}(t)$ in (\ref{Cell Width 2 t}) is the position of a fluid element at the instant $t$. These functions $t_{\text{m}}(z)$ or $z_{\text{m}}(t)$ and the values of $v_\text{N}$, $v_\text{CJ}$ and $D_\text{CJ}$ are determined by ZND numerical calculations with a detailed chemical kinetic mechanism (\hyperref[sec:Results]{Sect. 5} \hyperref[sec:App-ZND]{\& App. A}). The procedure for obtaining $\bar{\lambda}_\text{C}$ is simple post-processing of a ZND result spreadsheet, which amounts to finding the point where the ZND distance-time ratio is equal to the average material velocity (\ref{ZND Average Velocity}). That gives simultaneously $\bar{\ell}_{\text{Z}}$ and $\bar{t}_{\text{Z}}$, which respectively multiplied by $a\,k_1$ and $a\,k_2$ then give $\bar{\lambda}_\text{C}$.
From the relations (\ref{ZND Mass Conserv}), (\ref{ZND Average Volume}) and (\ref{Cell Length 2}), $\bar{y}$ also defines the mean burnt volume fraction $\bar{\nu}$ from the basic relations
    \begin{align}
    \bar{\ell}_{\text{ZI}}&=U_{\text{N}}\;\bar{t}_{\text{ZI}}=(1-\bar{\nu})\bar{\ell}_{\text{Z}} \label{MeanFracVol_I}\\
    \delta \bar{\ell}_{\text{ZR}}&=U_{\text{CJ}}\;\delta \bar{t}_{\text{ZR}}=\bar{\nu}\bar{\ell}_{\text{Z}} \label{MeanFracVol_R}\\
    \bar{\nu}&=\frac{\delta \ell_\text{ZR}}{\bar{\ell}_{\text{Z}}}=\bar{y}\;\frac{v_\text{CJ}}{\bar{v}_\text{Z}(\bar{y})}. \quad \label{MeanFracVol}
    \end{align}

Additionally, the model substantiates the proportionality relation (\ref{eq:K_lz}) between the cell mean width $\bar{\lambda}_\text{C}$, that the ZND characteristic length should include both thicknesses of the induction and main-reaction layers, \textit{i.e.} $\bar{\ell}_{\text{Z}}=\bar{\ell}_{\text{ZI}}+\delta\bar{\ell}_{\text{ZR}}$ (\hyperref[sec:mod-overview]{Sects. 2}\hyperref[sec:Results]{ \& 5}), and it gives the proportionality coefficient $k=a\,k_{2}$ as a function of the mean burnt mass fraction $\bar{y}$ (\hyperref[sec:GraphProba]{Sect. 4}) and the initial, shock and CJ specific volumes, respectively, $v_0$, $v_\text{N}$ and $v_\text{CJ}$.

\clearpage
\section{\label{sec:GraphProba}Graph theory and geometric probabilities}

\begin{figure}[ht!]
\centering
\includegraphics[height=0.9\linewidth]{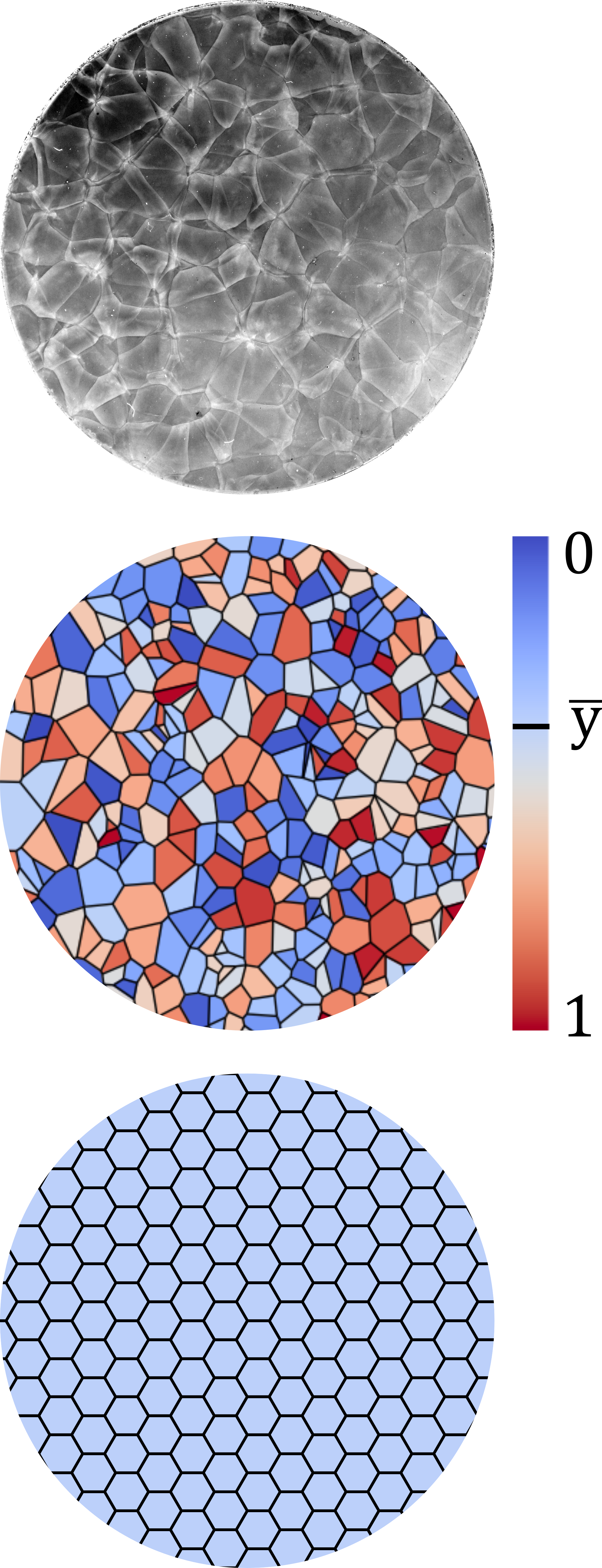}
\caption{Front views of the cellular structure. Top: soot recording for the \ce{2H2 + O2 +2Ar} mixture at $p_0=30$ kPa in a $16$ cm$^2$ round tube that illustrates the irregularity of the front-view patterns although the longitudinal are not. Middle: schematic with randomly distributed face colors representing randomly distributed burnt mass fraction $y(t)$. Bottom: equivalent hexagon-patterned tesselation with burnt mean mass fraction $\Bar{y}=0.38498$ (\ref{MeanFracMass}).}
\label{fig:model_steps}
\end{figure}

In \hyperref[sec:Introduction]{Section 1}, we have physically substantiated the premise of stationary ergodicity that the distributions of the geometric properties of a large number of irregular front-view patterns - the number of triangles, rectangles, pentagons, hexagons, \textit{etc.}, - should not be statistically different at one front position or another during the same experiment or at the same front position in one experiment or another. That was the principle of our preliminary analysis in \cite{Monnier2022a}. We pointed out that, in the limit of a high number of patterns $F$, elements from planar graph theory show that such front views are equivalent to tessellations of hexagons for pattern distribution independent of the front position. That was obtained by combining the physical condition that only three transverse waves can intersect with the mathematical limit at large $F$ of the Descartes-Euler-Poincar\'{e} relation $F-E+V=2$, which connects the number of faces $F$ (the cells), edges $E$ (the transverse waves) and vertices $V$ (the edge intersections) in a tessellation. For three-edge vertices, $2E=3V$, so the limit at large $F$ of the number of edges per face $2E/F$ is $6$, hence the hexagon (Fig. \ref{fig:model_steps}). Thus a cell counting on an experimental recording gives the estimate of the cell mean width
\begin{equation}
\bar{\lambda}_\text{C}=\frac{3\ln 3}{\pi }\frac{\sqrt{3}}{2}d_{\text{x}},\quad d_{%
\text{x}}=\sqrt{\frac{8}{3\sqrt{3}}A_{\text{C}}},\quad A_{\text{C}}=\frac{A_{%
\text{T}}}{F},  \label{Graph_Theo}
\end{equation}
where $d_{\text{x}}$ and $A_{\text{C}}$ are the outer diameter and the area of the hexagon, and $A_{\text{T}}$ the cross-section area of the tube.
In the present work, the important consequence of this premise is that a combination of properties of this tessellation of hexagons and geometric probabilities predicts the mean reaction progress variable $\bar{y}$ and hence the cell mean width $\bar{\lambda}_\text{C}$ (\hyperref[sec:Model]{Sect. 3}).

First, we define a control volume with the surface area $A_{\text{C}}$ and the half length $\bar{L}_{\text{C}}/2$ of the ideal cell (\ref{fig:ZND_Cell}). We denote by $M_{\text{C}}$ the mass contained in this volume, $M\left(t\right)$ the mass that has crossed this surface during the period $0\leqslant t\leqslant \bar{t}_{\text{C}}/2$, \textit{i.e.} when the front has travelled the distance $L\left(t\right)=Dt \leqslant \bar{L}_{\text{C}}/2$, and, respectively, $A_{\text{B}}\left(t\right)$ and $M_{\text{B}}\left(t\right)$ the area swept and the mass burnt by the transverse waves during this period (Fig. \ref{fig:ZND_Cell}-right). They write 
\begin{equation}
M_{\text{C}}=\rho _{0}A_{\text{C}} \frac{\bar{L}_{\text{C}}}{2},\quad
M\left( t\right)=\rho _{0}A_{\text{C}}L\left(t\right) ,\quad M_{%
\text{B}}\left(t\right)=\rho_{0}A_{\text{B}}\left(t\right)L\left(t\right), \label{Masses}
\end{equation}
where $\rho _{0}$ denotes the initial specific mass. Thus, the burnt mass fraction $y_{\text{C}}$ is
\begin{equation}
y_{\text{C}}\left(t\right) =\frac{M_{\text{B}}\left(t\right)}{M\left(t\right)}=\frac{A_{\text{B}}\left( t\right)}{A_{\text{C}}},
\end{equation}
so its mean $\bar{y}_{\text{C}}$ (\ref{Time Averages}) is the mean combustion area respective to the cell area $A_{\text{C}}$.

Next, we express the stationary ergodicity of the transverse waves viewed as a large set of line segments of different lengths. We assume the successive orientations and lengths of the transverse waves over the propagation period $\bar{t}_{\text{C}}/2$ in the same experiment to be statistically identical to those in one experiment or another at the same front position. That ensures that the combustion efficiency is, on average, independent of the experiment and the front position, which is tantamount to admitting that the transverse waves behave like line segments randomly dropped on the front surface. Thus, $\bar{y}_{\text{C}}$ is the probability that the segments are entirely contained in the cell surface, \textit{i.e.} the non-intersection probability. Its calculation is a classical problem of geometric probabilities, namely Buffon's needle problem extended to a surface with a polyhedral tiling and needle lengths varying between $0$ and the largest width of the polyhedron, here for example, the outer diameter $d_{\text{x}}$ of the hexagon.

Many accounts of such problems express a non-intersection probability as a ratio $\mu _{\text{C}}/\mu$ of measures. The measure $\mu$ is the hyper-volume of the space of all possible random values of the independent variables, and $\mu_{\text{C}}$ is that of the subspace of those values that ensure the non-intersection of the segments with the edges of the polyhedron. The independent variables are the orientation angle, the maximum length and the center coordinates of the segment, given the polyhedron. The non-intersection constraint confines the segment center to a smaller polyhedron whose shape and area depend on the segment orientation and length. For the hexagon, we extend below to a variable-length segment the solution of Vassallo \cite{Vassallo2021} for a constant-length segment. Because of the clarity of his presentation, we do not reproduce his calculations for brevity. In our notation, the hexagon has an area $A_{\text{C}}=(3\sqrt{3}/2)\times (d_{\text{x}}/2)^2$, the segment angle varies in $\left[ 0,2\pi \right]$ and its length $s$ in $\left[ 0,d_{\text{x}}\right]$, so the measure is $A_{\text{C}} \times 2\pi \times d_{\text{x}} = 6\pi \sqrt{3} (d_{\text{x}}/2)^3$. Nondimensionalizing the lengths $s$ by the side length $d_{\text{x}}/2$ of the hexagon, and denoting by $r=2s/d_{\text{x}}$ the non-dimensional segment lengths, we obtain
\begin{eqnarray}
\mu &=&6\pi \sqrt{3},\quad \mu _{\text{C}}=\mu_{1}+\mu_{2}+\mu_{3},\quad
\mu_{i}=\int_{r_{i1}}^{r_{i2}}m _{i}\left( r\right) dr, \\
r &\in &\left[ r_{11}=0,r_{12}=1\right] ,\quad m_{1}\left( r\right)=3\pi 
\sqrt{3}-12r+r^{2}\left( 3-\pi /\sqrt{3}\right) , \\
r &\in &\left[ r_{21}=1,r_{22}=\sqrt{3}\right] ,\quad m_{2}\left(
r\right) =\pi \sqrt{3}\left( r^{2}+5\right) -9\sqrt{4r^{2}-3}\;... \nonumber\\
&&\quad\quad...\;-2\sqrt{3}\left( 3+2r^{2}\right) \arcsin \left(\sqrt{3}/2r\right) , \\
r &\in &\left[ r_{31}=\sqrt{3},r_{32}=2\right] ,\quad m_{3}\left(
r\right) =2\sqrt{3}\left( r^{2}+12\right) \arcsin \left(\sqrt{3}/r\right) \;... \nonumber\\
&&\quad\quad...\;+30\sqrt{r^{2}-3}-\left( 8\pi \sqrt{3}+18\right) - r^{2}%
\left(3+2\pi/\sqrt{3}\right) , \\
\mu_{1} &=&10.720,\quad \mu_{2}=1.\,837\,4,\quad \mu_{3}=1.154\,7\times 10^{-2},
\end{eqnarray}
where the $m_{i}$\hspace{0.025cm}s are Vassallo's non-intersection measures for constant segment lengths \cite{Vassallo2021} and the $\mu_{i}$\hspace{0.025cm}s ours for segment lengths varying in the intervals $[ 0,d_{\text{x}}/2]$, $[d_{\text{x}}/2,\sqrt{3}d_{\text{x}}/2]$ and $[\sqrt{3}d_{\text{x}}/2,d_{\text{x}}]$. This gives the non-intersection probability, that is, the mean burnt mass fraction $\bar{y}\equiv \bar{y}_{\text{C}}$ (\hyperref[sec:Model]{Sect.3}), by
\begin{equation}
\mu =32.648,\quad \mu_{\text{C}}=12.569,\quad \bar{y}=\frac{\mu_{\text{C}}}{\mu}\approx 0.38498. \label{MeanFracMass}
\end{equation}

Finally, we obtain the aspect ratio $\bar{\lambda}_\text{C}/\bar{L}_{\text{C}}$ by combining stochasticity and, inspired by \cite{Takai1975}, geometry. Since the transverse waves have a stochastic motion, their positions can be considered to be the same every period $\bar{t}_{\text{C}}$, so the longitudinal overdriven front waves of the model cellular front should superimpose on each other every distance $\bar{L}_{\text{C}}$. Equivalently, these waves can be viewed as the upper surface elements of spheres arranged in the hexagonal closest packing, that is, with alternate layers in the ABAB ... sequence. The sphere diameter is also the distance between the centers of adjacent spheres and the inner diameter $d_{\text{i}}$ of a hexagon, so the ratio $\bar{L}_{\text{C}}/d_{\text{i}}$ comes out as twice the height of the tetrahedral pyramid whose base is the triangle with vertices the centers of the three closest spheres in the same layer. Simple geometry then gives $\bar{L}_{\text{C}}/d_{\text{i}}=\sqrt{8/3}$ and $d_{\text{i}}/d_{\text{x}}=\sqrt{3}/2$. With the first relation (\ref{Graph_Theo}), that yields the mean cell aspect ratio
\begin{equation}
a=\frac{\bar{\lambda}_\text{C}}{\bar{L}_{\text{C}}}=\frac{3\ln 3}{\pi }\sqrt{\frac{3}{8}}%
\approx 0.64244, \label{AspRatio}
\end{equation}
which gives the opening angle $65.4^\text{o}$ well representing the measurements on the longitudinal recordings.
These values $\bar{y}=0.38498$ and $a=0.64244$ close the compatibility system that defines the value of $\bar{\lambda}_{\text{C}}$ from the intersection of $\lambda_{\text{1}}(z)$ (\ref{Cell Width 1 z}) and $\lambda_{\text{2}}(z)$ (\ref{Cell Width 2 z})

This statistical description is independent of whether the cellular mechanism is adiabatic. Thus, why the mean burnt mass fraction $\bar{y}$ and thus the time ratios $\bar{t}_{\text{C}}/\bar{t}_{\text{Z}}$, $\delta t_{\text{ZR}}/\bar{t}_{\text{Z}}$ and $\bar{t}_{\text{ZI}}/\bar{t}_{\text{Z}}$ (\hyperref[sec:Model]{Sect. 3}) are pure numbers, independent of the chemical and physical properties of the mixture, follows from the combination of the model hypothesis that the transverse wave motion is a stationary ergodic process with the limit at large cell number of the Descartes-Euler-Poincar\'{e} relation. Indeed, this relation and its limit are independent of time. The value $\bar{y}\approx0.385$ indicates that the crossing times of the induction and main-reaction layers are the fractions $61.5\%$ and $38.5\%$, respectively, of the representative ZND complete reaction time $\bar{t}_\text{Z}$, and that the model adiabatic cellular process is $\bar{t}_{\text{C}}/\bar{t}_{\text{Z}}\approx5.2$ longer than the ZND process, \textit{i.e.} takes $\approx5.2$ longer to achieve complete reaction. In contrast, from (\ref{MeanFracVol_I})-(\ref{MeanFracVol}), the mean burnt volume fraction $\bar{\nu}$ and thus the length ratios $L_{\text{C}}/\bar{\ell}_{\text{Z}}$, $\bar{\ell}_{\text{ZI}}/\bar{\ell}_{\text{Z}}$ and $\delta \bar{\ell}_{\text{ZR}}/\bar{\ell}_{\text{Z}}$ depend on the shock and reaction-end properties. Table \ref{Tab_Check} in \hyperref[sec:Results]{Section 5} shows an example.

\clearpage
\section{\label{sec:Results}Results}

\begin{figure*}[ht!]
\centering
\begin{subfigure}[b]{0.5\textwidth}
  \centering
  \includegraphics[width=1\linewidth]{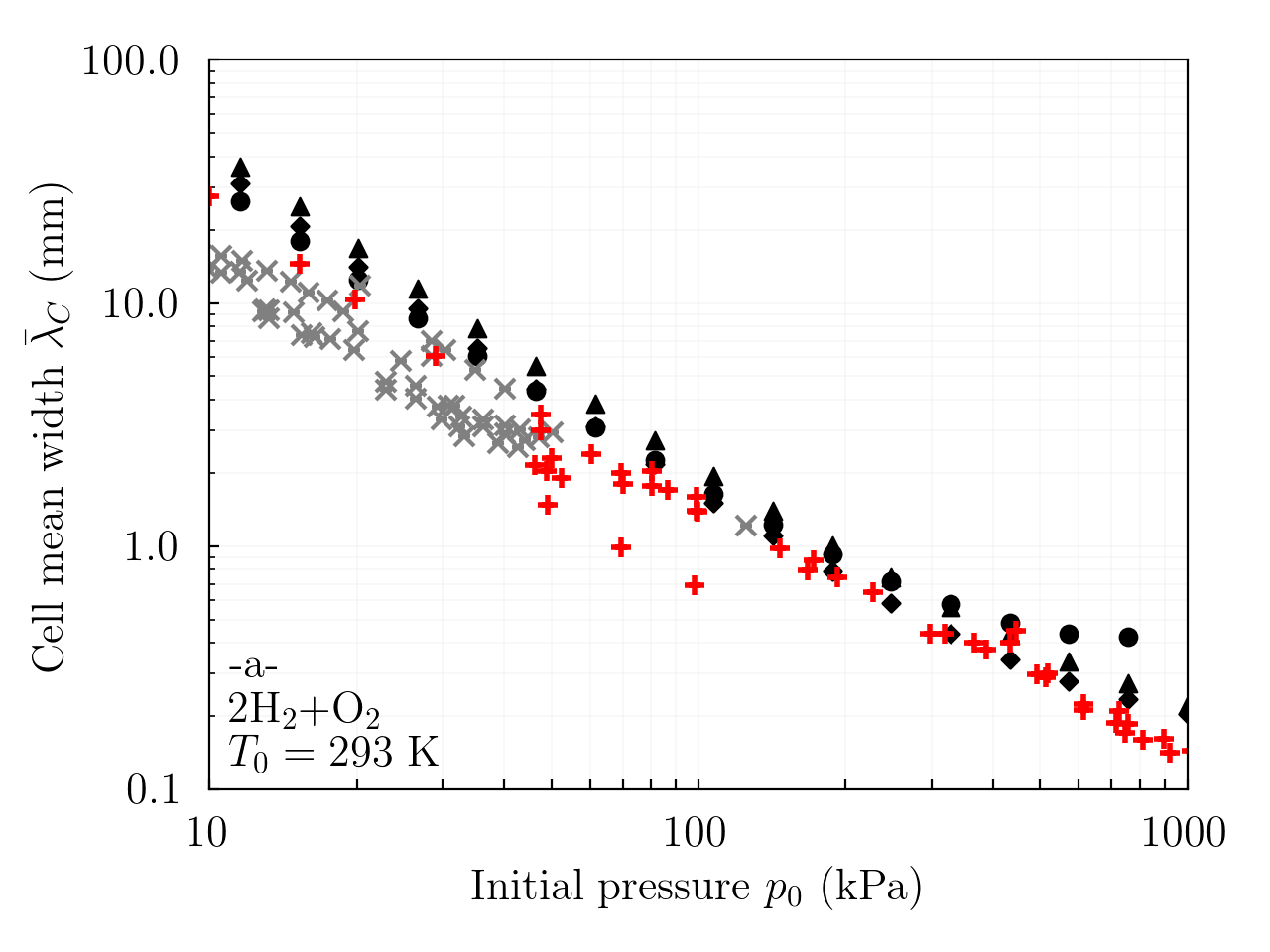}
\end{subfigure}
\hspace{-0.5cm}
\begin{subfigure}[b]{0.5\textwidth}
  \centering
  \includegraphics[width=1\linewidth]{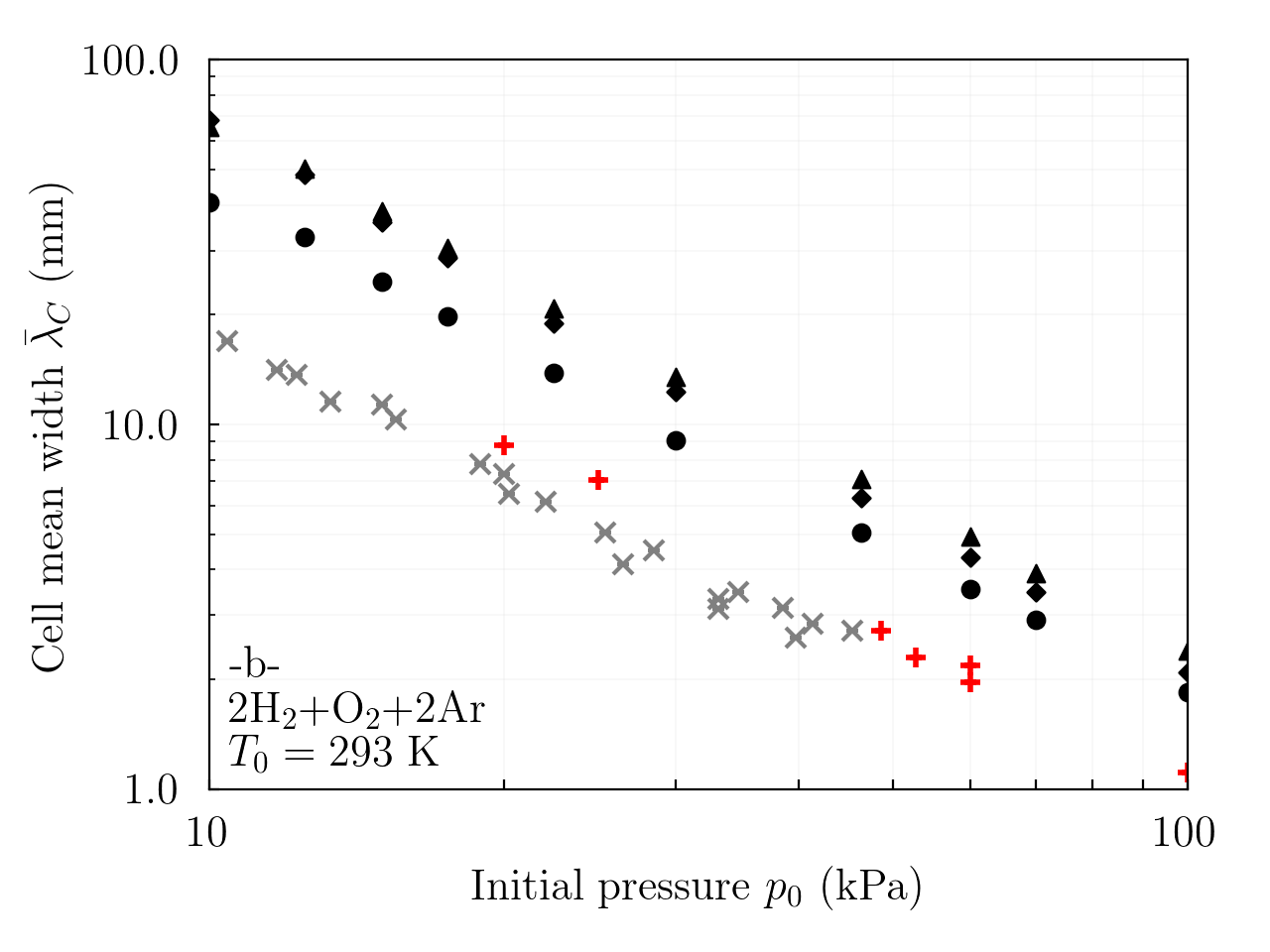}
\end{subfigure}
\centering
\begin{subfigure}[b]{0.5\textwidth}
  \centering
  \includegraphics[width=1\linewidth]{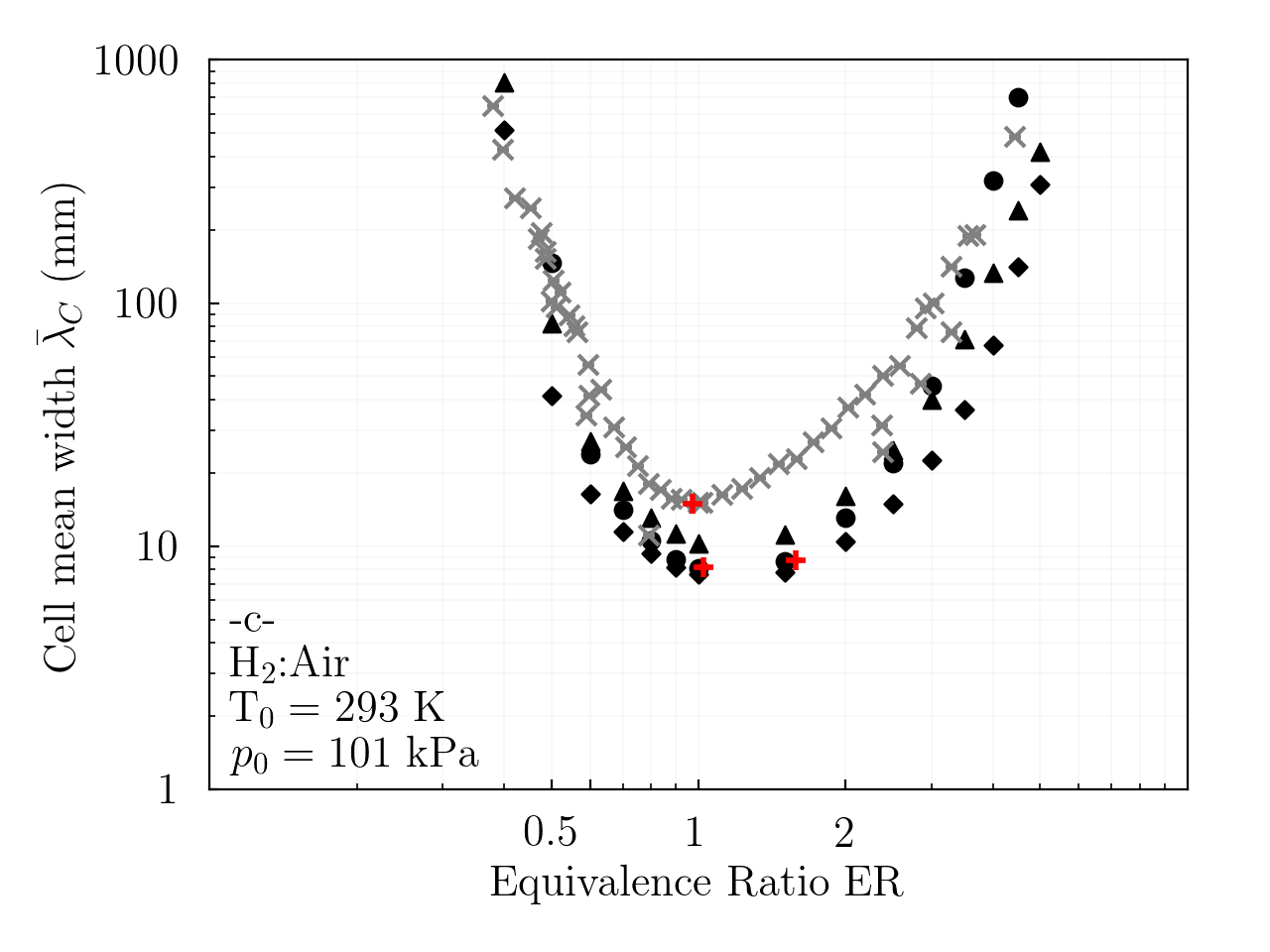}
\end{subfigure}
\hspace{-0.5cm}
\begin{subfigure}[b]{0.5\textwidth}
  \centering 
  \includegraphics[width=1\linewidth]{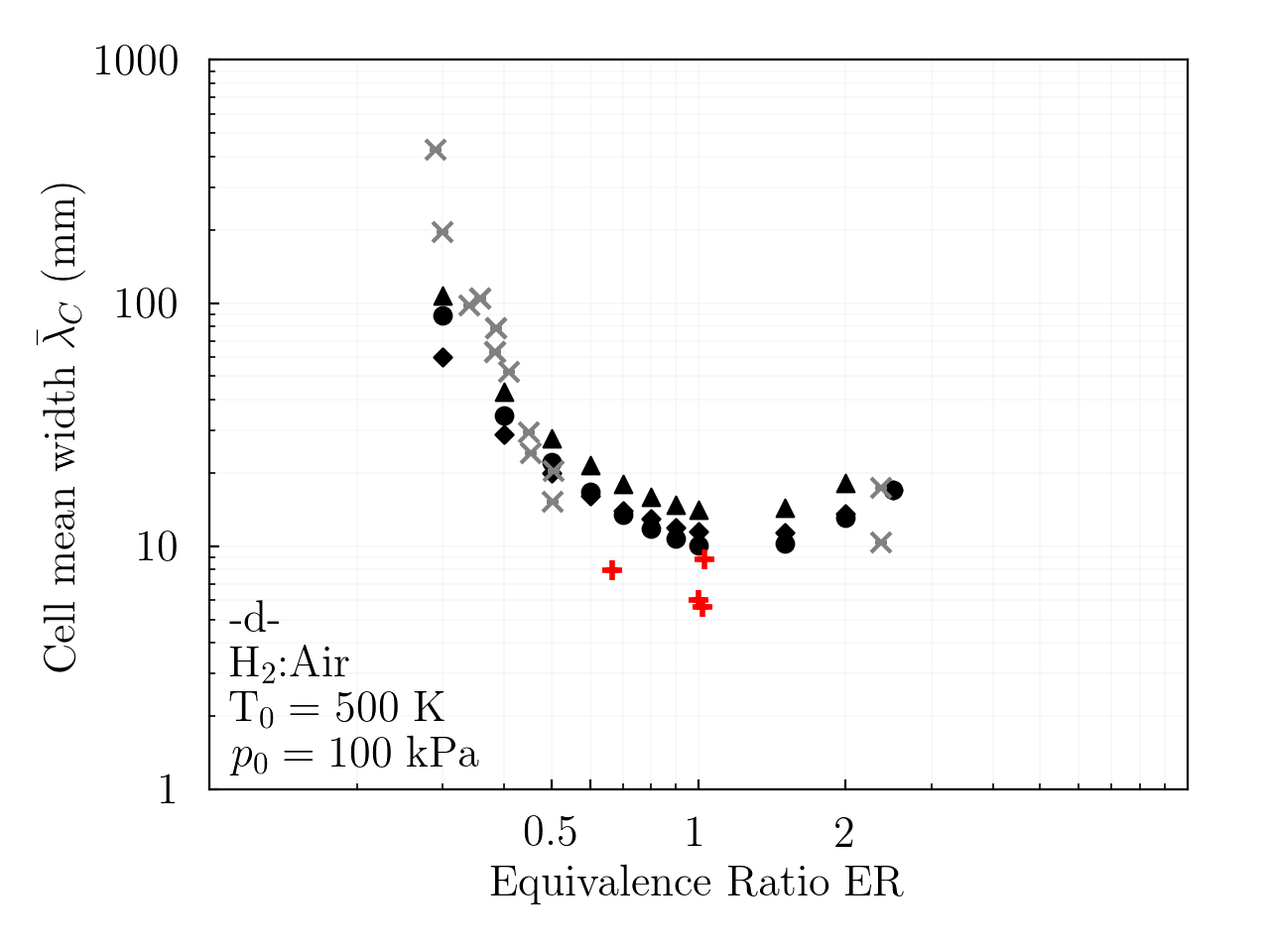}
\end{subfigure}
\centering
\begin{subfigure}[b]{0.5\textwidth}
  \centering  
  \includegraphics[width=1\linewidth]{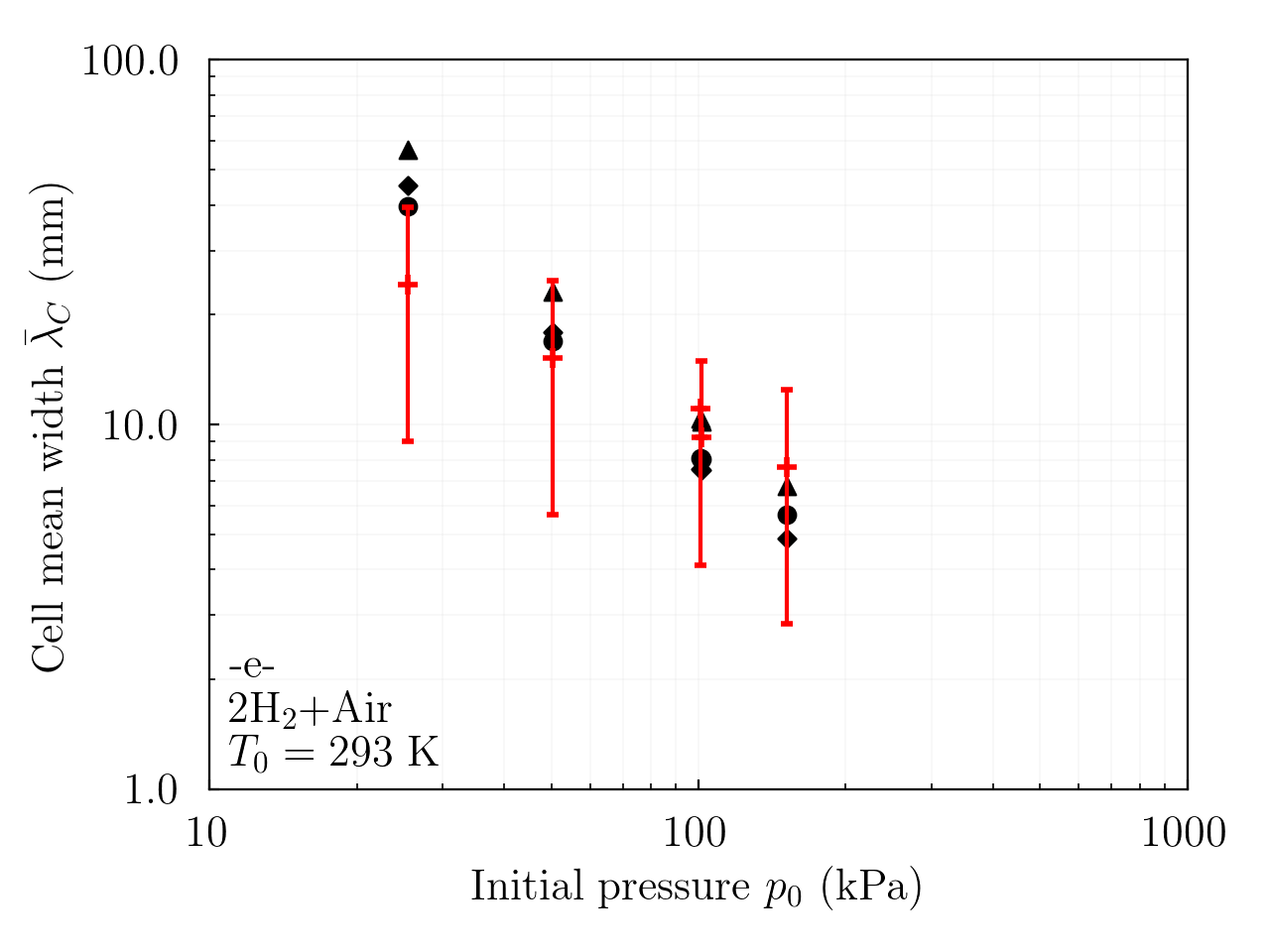}
\end{subfigure}
\hspace{-0.5cm}
\begin{subfigure}[b]{0.5\textwidth}
  \centering
  \includegraphics[width=1\linewidth]{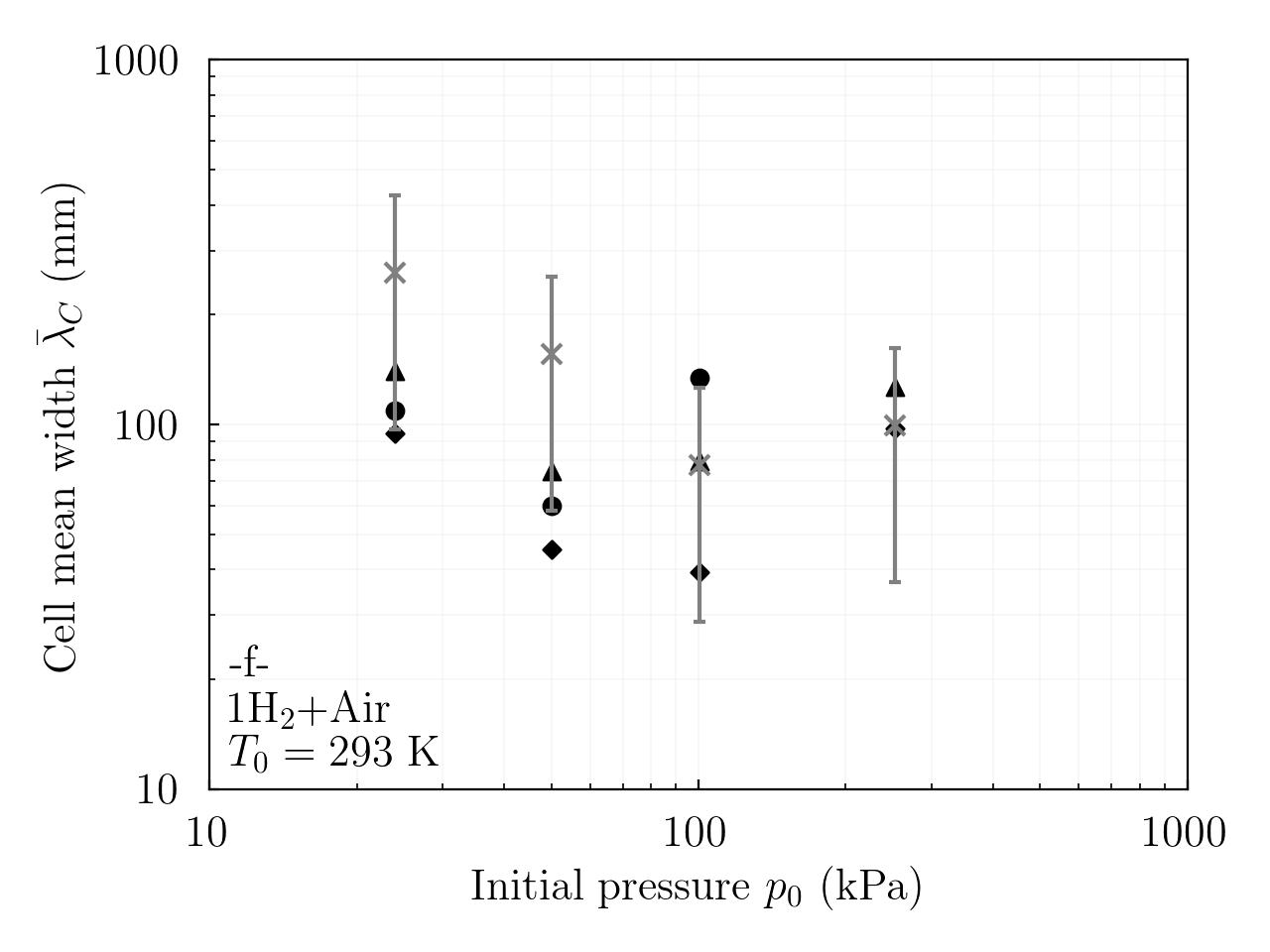}
\end{subfigure}
\vspace{0.5cm}
\caption{
Calculated and measured cell mean widths $\bar{\lambda}_\text{C}$ for the \ce{H2} fuel. Full and open symbols: calculations using Konnov's \cite{Konnov2007} ($\protect\markerkonnovB$), the San Diego \cite{SanDiego} ($\protect\markersandiegoB$) and the FFCM-1 \cite{Smith2016} ($\protect\markerffcmB$) mechanisms of chemical kinetics. Crosses: measurements \cite{DATABASE} with large ($\geqslant\mathcal{O}(10)$, red {\scriptsize${\color{red}{\bm{+}}}$}) and small ($\leqslant\mathcal{O}(10)$, gray {\scriptsize${\color{gray}{\bm{\times}}}$}) ratios $d_\text{transv}/\bar{\lambda}_\text{C}$, with $d_\text{transv}$ the transverse dimension of the tubes.
\label{fig:Graphs_H2}
}
\end{figure*}

\begin{figure*}[ht!]
\centering
\begin{subfigure}{0.5\textwidth}
  \centering
  \includegraphics[width=1\linewidth]{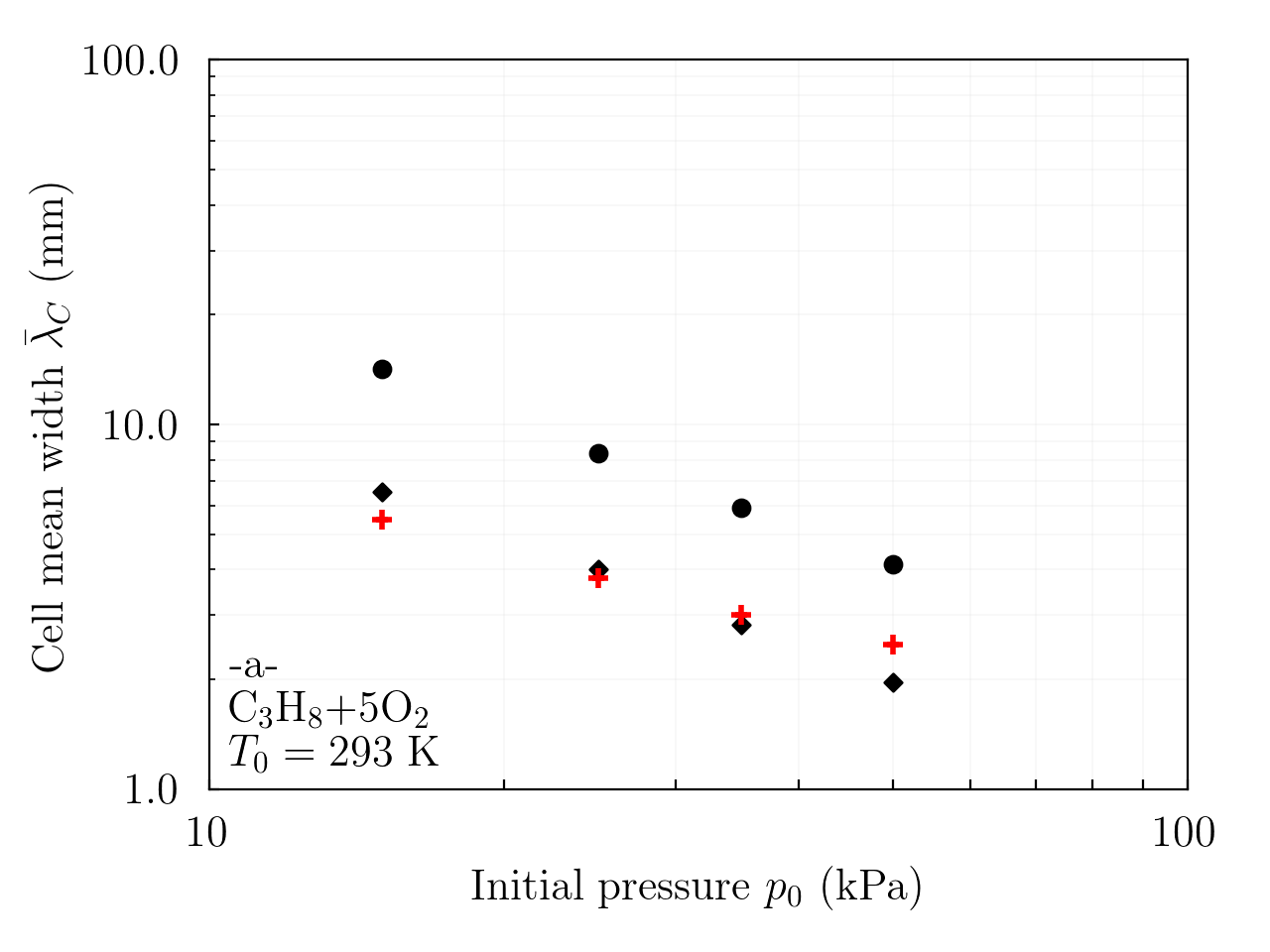}
\end{subfigure}
\hspace{-0.5cm}
\begin{subfigure}{0.5\textwidth}
  \centering
  \includegraphics[width=1\linewidth]{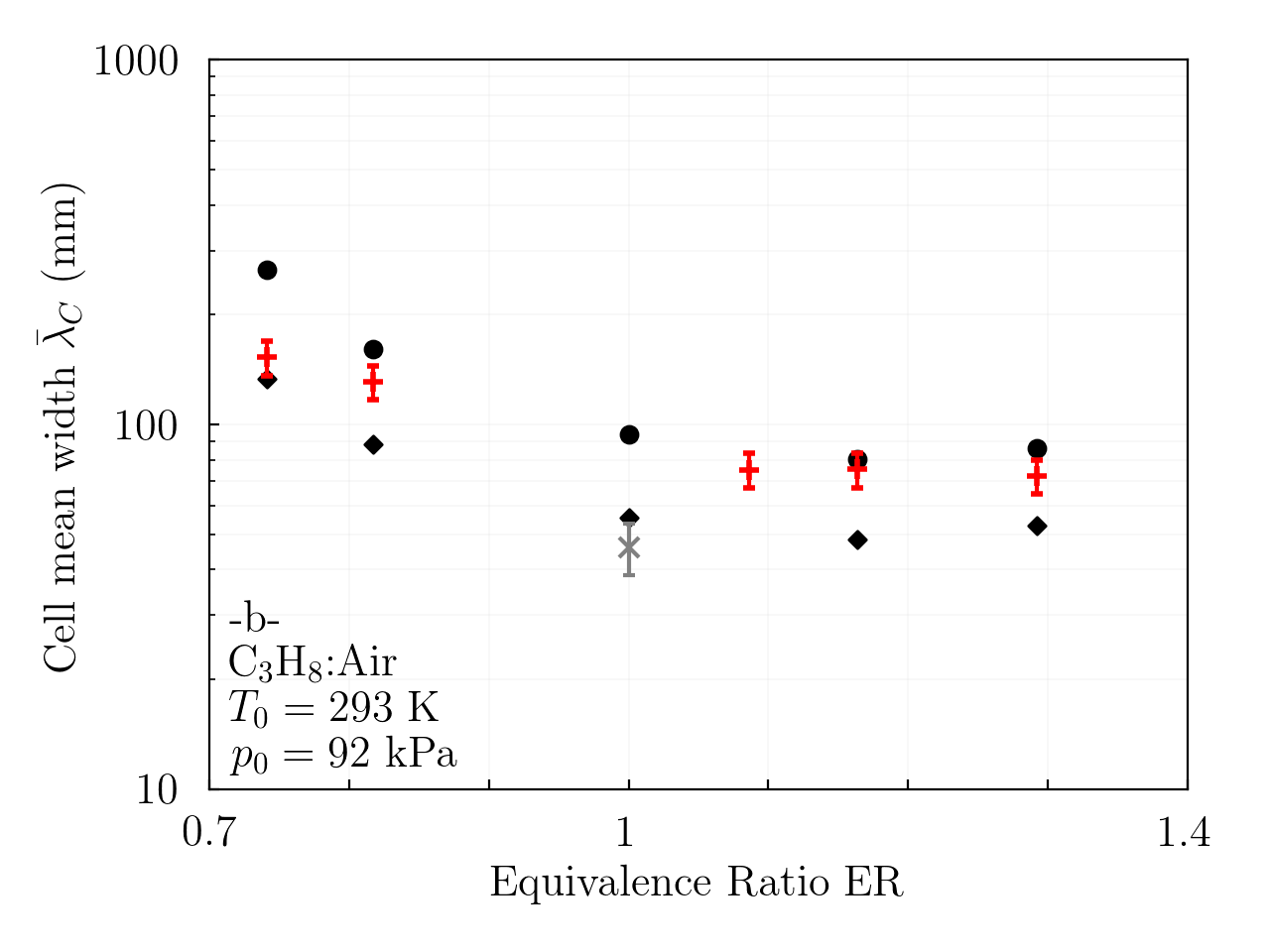}
\end{subfigure}
\begin{subfigure}{0.5\textwidth}
  \centering
  \includegraphics[width=1\linewidth]{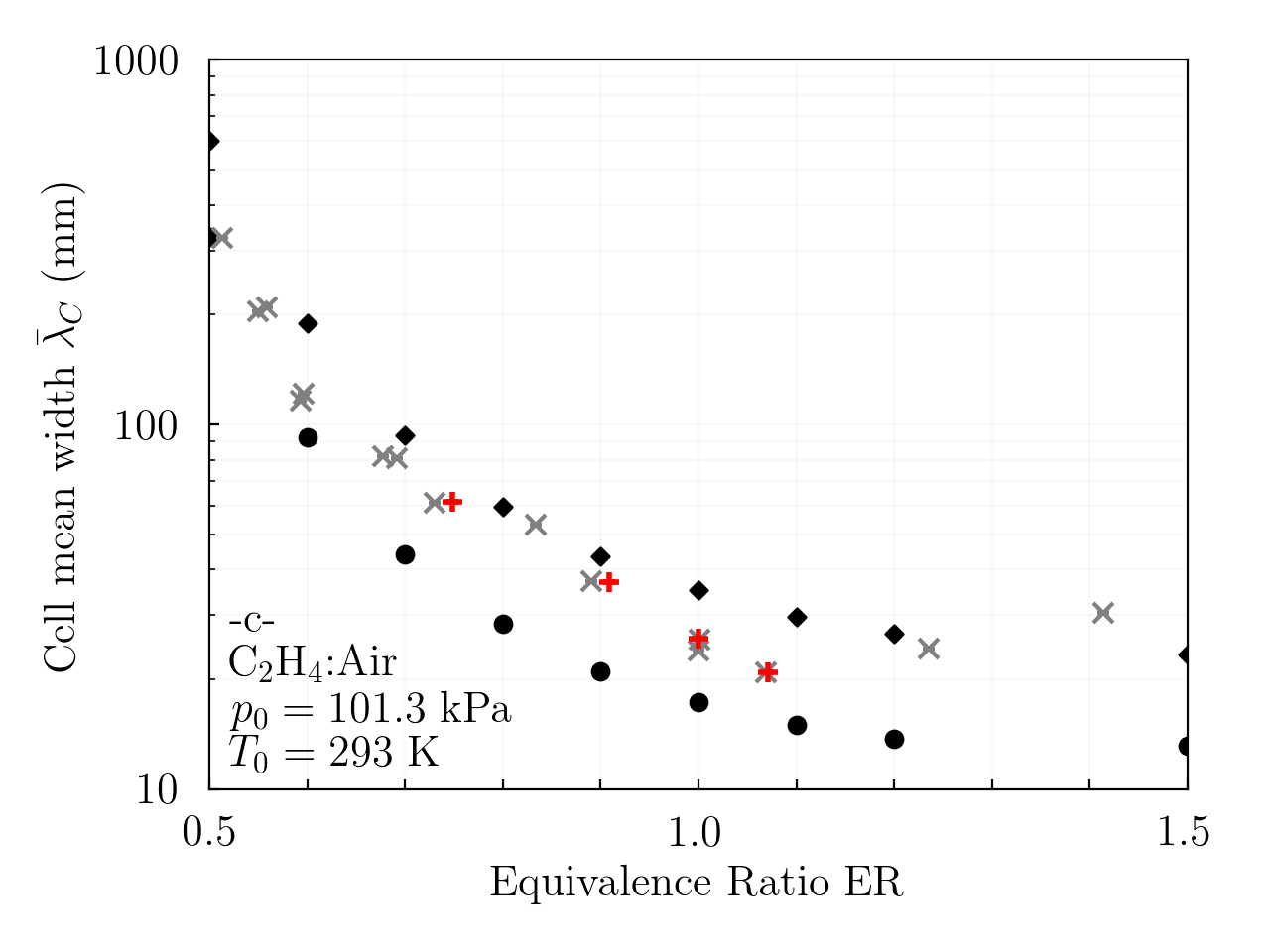}
\end{subfigure}
\hspace{-0.5cm}
\begin{subfigure}{0.5\textwidth}
  \centering
  \includegraphics[width=1\linewidth]{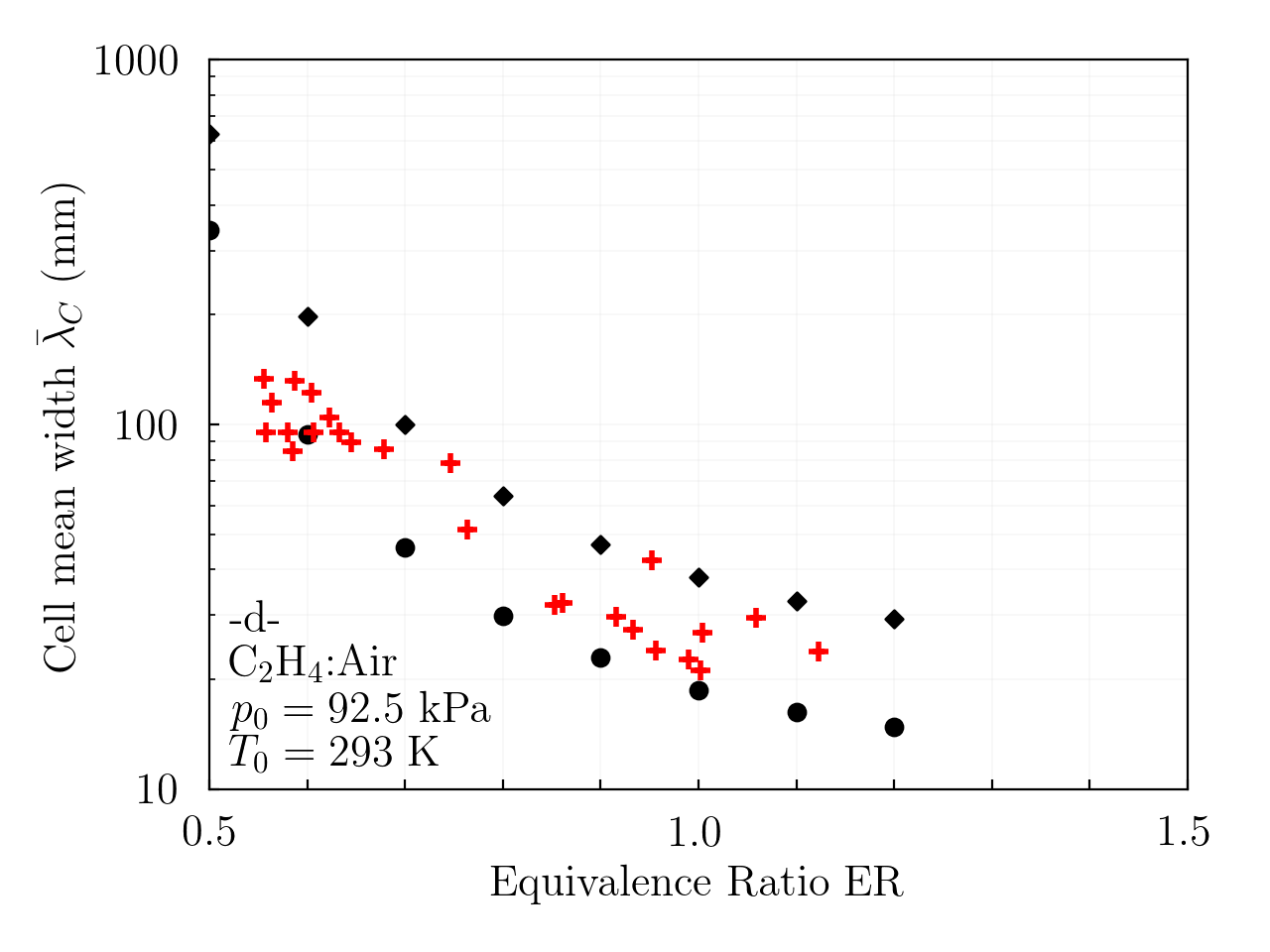}
\end{subfigure}
\begin{subfigure}{0.5\textwidth}
  \centering
  \includegraphics[width=1\linewidth]{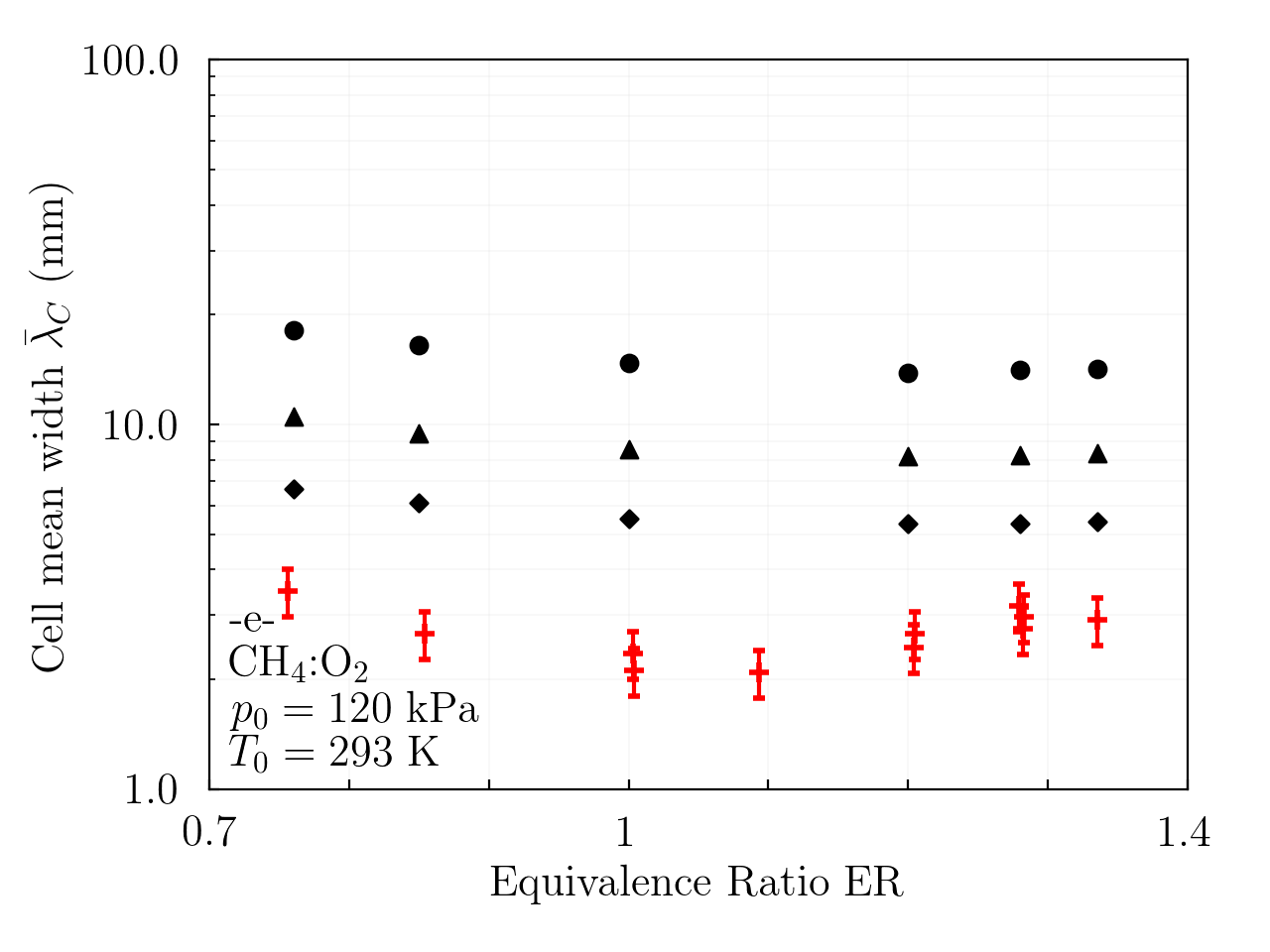}
\end{subfigure}
\hspace{-0.5cm}
\begin{subfigure}{0.5\textwidth}
  \centering
  \includegraphics[width=1\linewidth]{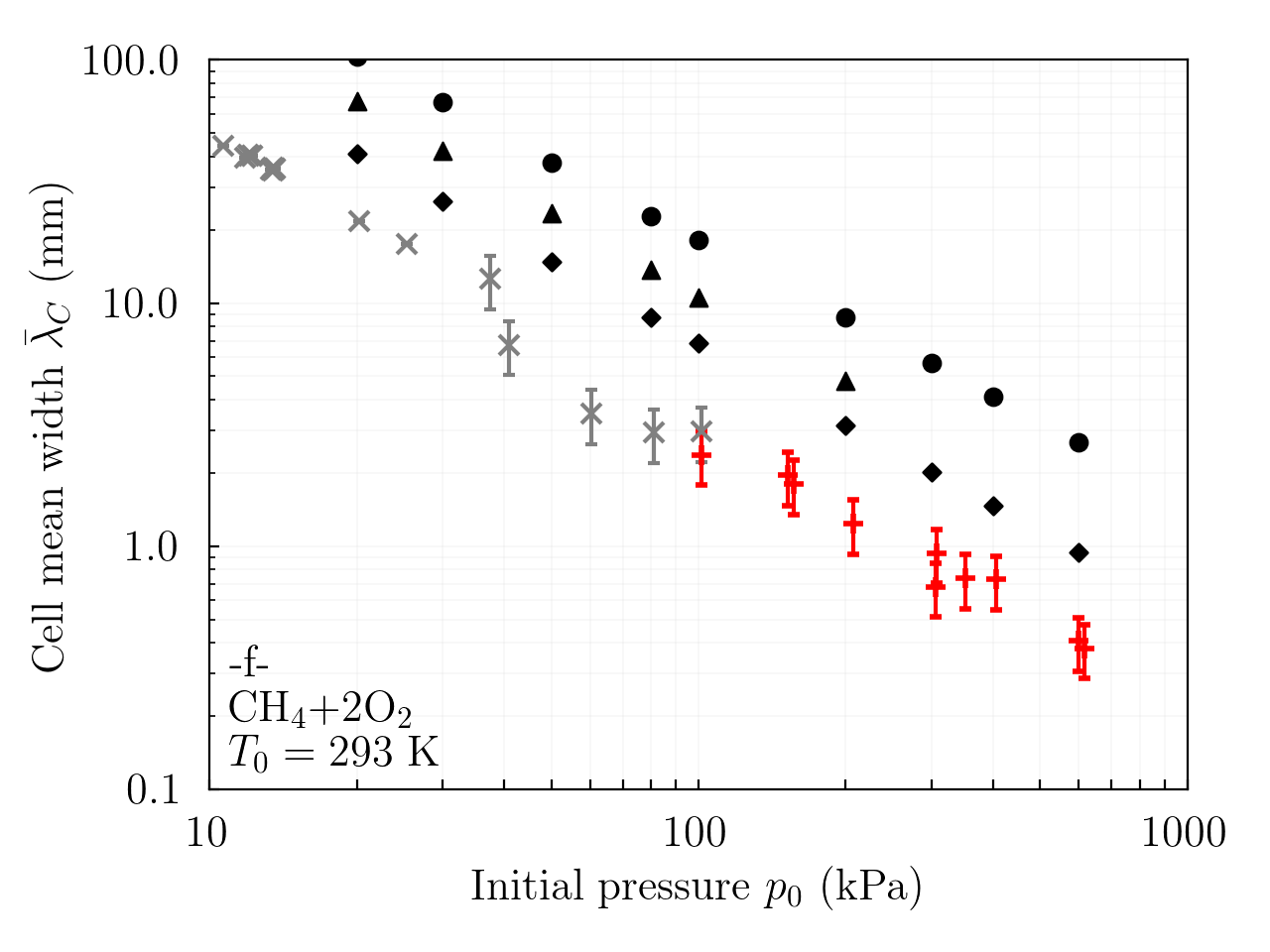}
\end{subfigure}
\vspace{0.5cm}
\caption
{Calculated and measured cell mean widths $\bar{\lambda}_\text{C}$ for hydrocarbon fuels. Full and open symbols: calculations using Konnov's \cite{Konnov2007} ($\protect\markerkonnovB$), the San Diego \cite{SanDiego} ($\protect\markersandiegoB$) and the FFCM-1 \cite{Smith2016} ($\protect\markerffcmB$) mechanisms of chemical kinetics. Crosses: measurements \cite{DATABASE} with large ($\geqslant\mathcal{O}(10)$, red {\scriptsize${\color{red}{\bm{+}}}$}) and small ($\leqslant\mathcal{O}(10)$, gray {\scriptsize${\color{gray}{\bm{\times}}}$}) ratios $d_\text{transv}/\bar{\lambda}_\text{C}$, with $d_\text{transv}$ the transverse dimension of the tubes.
}
\label{fig:Graphs_CnHm}
\end{figure*}

Figures \ref{fig:Graphs_H2} and \ref{fig:Graphs_CnHm} compare calculated and experimental cell mean widths $\bar{\lambda}_\text{C}$. We have used the experimental values of the Caltech Detonation Database (DDB) \cite{DATABASE}.
We recall that comparison is valid only for a sufficiently large number of cells on the front surface (\hyperref[sec:Model]{Sects. 3}\hyperref[sec:GraphProba]{ \& 4}). Therefore, we have selected the measurements in tubes whose transverse dimension is at least ten times the experimental $\bar{\lambda}_\text{C}$, so there should be at least $\mathcal{O}(100)$ cells on the front. These measurements are indicated by the red crosses ({\scriptsize${\color{red}{\bm{+}}}$}). The measurements in too narrow tubes are shown by the gray crosses ({\scriptsize${\color{gray}{\bm{\times}}}$}).
This precaution shows that the model works better with large numbers of cells, \textit{i.e.} when the tube geometry has less influence on the cellular mechanism. 
The $\mathcal{O}(100)$ order of magnitude is arbitrary and follows from our experimental observations for the mixture \ce{2H2 + O2 + 2Ar} \cite{Monnier2022a,Monnier2022b}. Information on the tube dimensions and measurement uncertainties (shown as error bars in \ref{fig:Graphs_H2}-e) can be found in the references given in the DDB.
We carried out the ZND calculations \hyperref[sec:App-ZND]{(App. A)} using the Konnov's \cite{Konnov2007}, San-Diego \cite{SanDiego}, and FFCM-1 \cite{Smith2016} chemical kinetic mechanisms. We chose \ce{H2}, \ce{C3H8}, \ce{C2H4}, or \ce{CH4} as fuels and \ce{O2} as the oxidant, pure or diluted with the diatomic diluent \ce{N2} in its proportion in air (\ce{O2 + 3.71 N2 + 0.048 Ar}), because of their practical importance and the attention they have received from kineticists. We input the same initial pressures and temperatures as in the DDB, and the table \ref{Tab_mechanisms} shows which mechanism fits which fuel. 

\begin{table}[h!]
 \centering
 \caption{Compatibility table of the fuels and the chemical kinetic mechanisms} \label{Tab_mechanisms}
 \begin{tabular}[h!]{|P{2.6cm} | P{1.25cm} | P{1cm} P{1cm} P{1cm} P{1cm} | }
 \hline
 Fuel & symbol & \ce{H2} & \ce{C3H8} & \ce{C2H4} &\ce{CH4}  \\ [0.5ex] 
 \hline
 Konnov \cite{Konnov2007} & \markerkonnovB & yes & yes & yes & yes \\ [0.5ex] 
 \hline
 San-Diego \cite{SanDiego} & \markersandiegoB & yes & yes & yes & yes  \\ [0.5ex]
 \hline  
 FFCM-1 \cite{Smith2016} & \markerffcmB & yes & no & no & yes  \\
\hline
\end{tabular}
\end{table}

The model assumptions and the kinetic mechanism are the independent elements determining the quality of the calculated $\bar{\lambda}_\text{C}$. Their respective contributions to the differences with the experimental values are difficult to distinguish due to the sensitivity of the detonation characteristic lengths to the chemical kinetics.

Our approach to best separate them is to consider the mixture \ce{2H2 + O2}
as a reference case because its cell widths and chemical kinetic have been characterized sufficiently well for a long time. The good agreement shown in the figure \ref{fig:Graphs_H2}-a for each mechanism tends to confirm the validity of the model assumptions. Figures \ref{fig:Graphs_H2}-c to -e show that this favorable comparison extends to the mixtures \ce{H2}:\ce{Air} mixtures. In contrast, the figure \ref{fig:Graphs_H2}-b shows that the calculated $\bar{\lambda}_{\text{C}}$ of the mixture \ce{2H2 + O2 + 2 Ar}, \textit{i.e.} the mixture \ce{2H2 + O2} diluted with the large amount 40\% mol. of the monatomic diluent \ce{Ar} - are too high, even though the decreasing trend with increasing $p_0$ is correct. Since the model assumptions are valid for \ce{2H2 + O2} and that \ce{2H2 + O2 + 2 Ar} is considered more regular, a possible explanation is that the \ce{H2} sub-mechanism(s) in each of the three considered mechanisms is calibrated only for lean mixtures in case of large dilutions with \ce{Ar}, \textit{e.g.}, \cite{Konnov2019}. That contributes to emphasizing the sensitivity to chemical kinetics and the need for robust and physical mechanisms.

The figures \ref{fig:Graphs_CnHm}-a to -d for the hydrocarbon fuels \ce{C3H8} and \ce{C2H4}
also show as good an agreement as for \ce{H2}, considering the accepted experimental uncertainties, despite a larger sensitivity to the kinetic mechanism.
In contrast, the figures \ref{fig:Graphs_CnHm}-c and -f show that the calculations for \ce{CH4}:\ce{O2} mixtures overestimate the experimental $\bar{\lambda}_{\text{C}}$ regardless of the kinetic mechanism, while each is deemed valid in the investigated ranges of equivalence ratios and initial pressures. Our interpretation involves the too-high irregularity of such mixtures (\hyperref[sec:Introduction]{Sects. 1} \hyperref[sec:mod-overview]{ \& 2}). Firstly, the standard deviation of the measured cell widths for such mixtures is as large as the mean value, so a cell mean width is not a relevant characteristic length. Secondly, highly irregular cellular reaction zones are not driven by the shock-induced adiabatic cellular mechanism, so an agreement based on adiabatic modelling, such as ours, would not be relevant. Indeed, this mixture has the largest value of the $\chi$ parameter \hyperref[sec:Introduction]{(Sect. 1)}. Based on the expression $\chi=(E_\text{a}/RT_\text{N})(\bar{\ell}_\text{ZI}/\delta\bar{\ell}_\text{ZR})$, where $E_\text{a}$ denotes the global activation energy governing the induction zone and $R$ the universal gas constant, Ng and Zhang \cite{Ng2012_Zhang} give $\chi=52.5$ for \ce{CH4 + 2O2}, $\chi=16.6$ for \ce{C3H8 + 5 O2} and $\chi=2.25$ for \ce{2H2 + O2} at $p_0 = 0.2$ atm.

\begin{figure}[ht!]
\centering
\begin{subfigure}{0.455\textwidth}
    \centering
    \includegraphics[width=\linewidth]{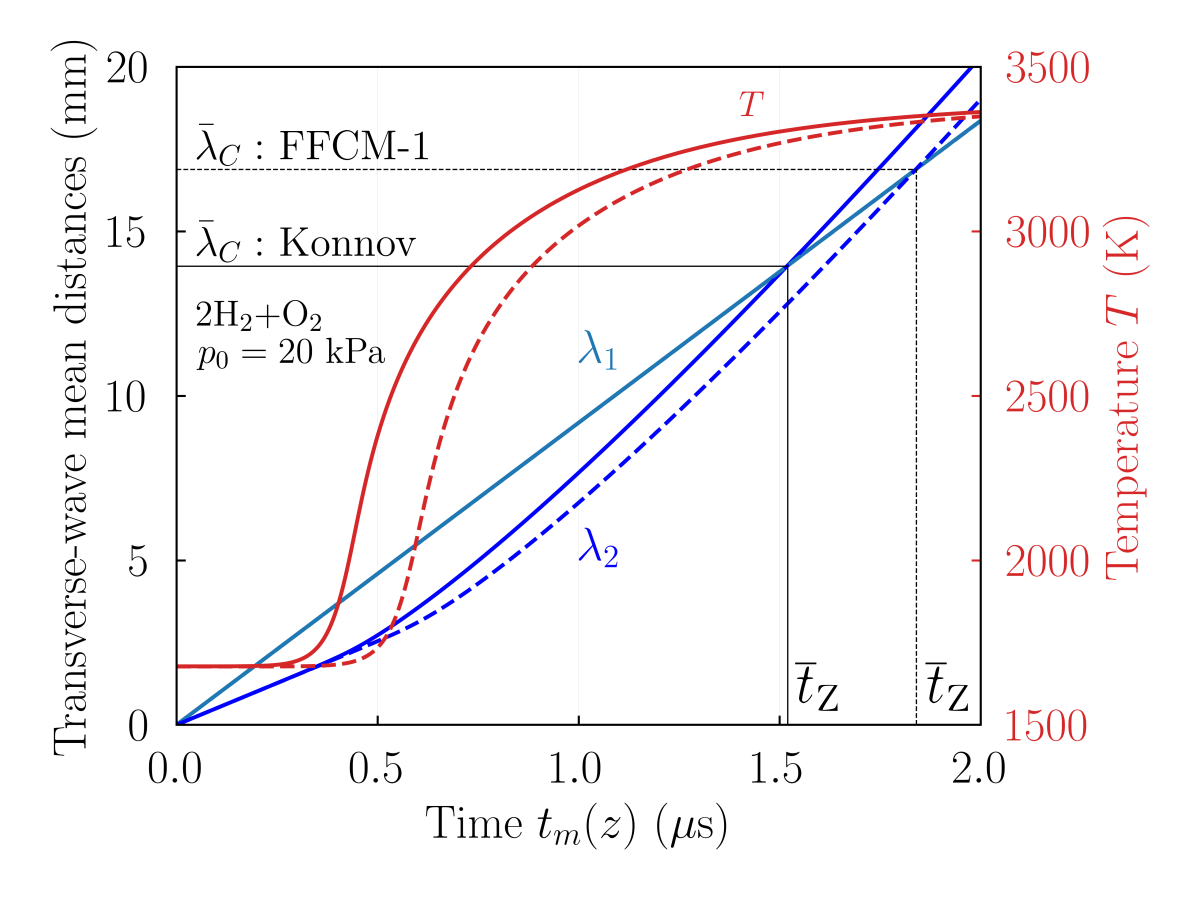}
\end{subfigure}
\begin{subfigure}{0.455\textwidth}
    \centering
    \includegraphics[width=\linewidth]{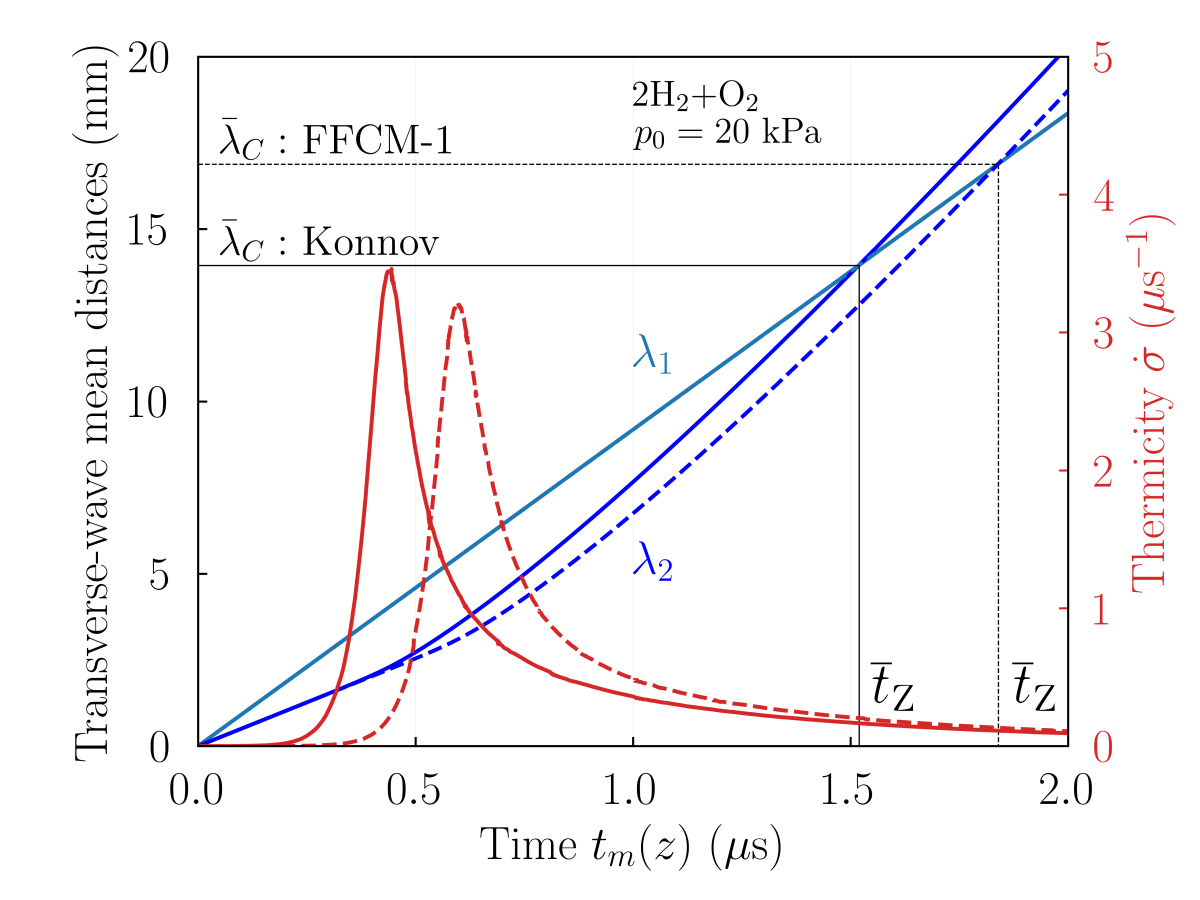}
    \hspace{-0.5cm}
\end{subfigure}
\caption{Sensitivities of the ZND temperature $T$ and thermicity $\Dot{\sigma}$ evolutions (red curves, right ordinate), the transverse-wave mean distances ${\lambda}_1$ and ${\lambda}_2$ (blue and green curves, left ordinate), the cell mean width $\bar{\lambda}_\text{C}$ and the ZND complete reaction time $\bar{t}_\text{Z}$ to the kinetic mechanism (full curves: Konnov's, dashed curves: FFCM-1). The green curves for each mechanism are indistinguishable. The intersections of ${\lambda}_1$ and ${\lambda}_2$ give the cell mean width $\bar{\lambda}_\text{C}$ and the ZND complete reaction time $\bar{t}_\text{Z}$ (\hyperref[sec:Model]{Sects. 3}\hyperref[sec:GraphProba]{ \& 4}).}
\label{fig:ZND_Konnov_FFCM1}
\end{figure}

Figure \ref{fig:ZND_Konnov_FFCM1} illustrates the sensitivity of the calculated cell mean width $\bar{\lambda}_{\text{C}}$ and the ZND evolutions of temperature $T$ and thermicity $\dot{\sigma}$ to the kinetic mechanism. The figure compares the curves obtained with the Konnov's and FFCM-1 mechanisms for the reference mixture \ce{2H2 + O2}. Indeed, this is the case for which Figure \ref{fig:Graphs_H2}-a shows results indistinguishable from one mechanism to the other. The log-log representation best reflects the agreement over a wide range of initial pressures but precludes the assessment of differences at a given initial pressure. Figure \ref{fig:ZND_Konnov_FFCM1} shows that all relative differences are small. In particular, that of the mean cell width $\bar{\lambda}_{\text{C}}$ is smaller than the accepted experimental uncertainty of $10-25\%$ \cite{Manzhalei1974,Bull1982,Moen1984,Tieszen1986,Aminallah1993}.

\begin{figure*}[ht!]
\centering
\begin{subfigure}{0.4\textwidth}
  \centering
  \includegraphics[width=1\linewidth]{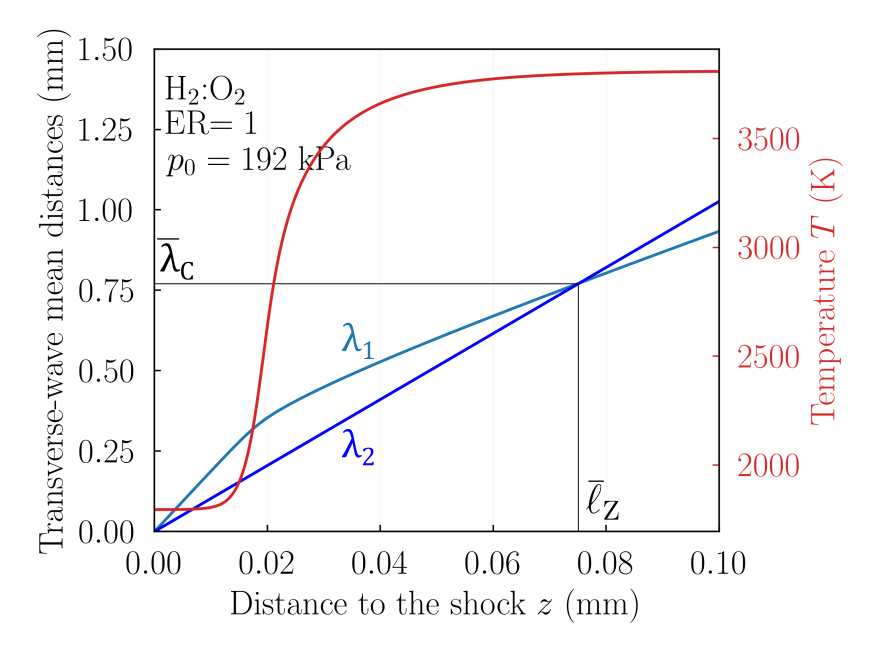}
\end{subfigure}
\begin{subfigure}{0.4\textwidth}
  \centering
  \includegraphics[width=1\linewidth]{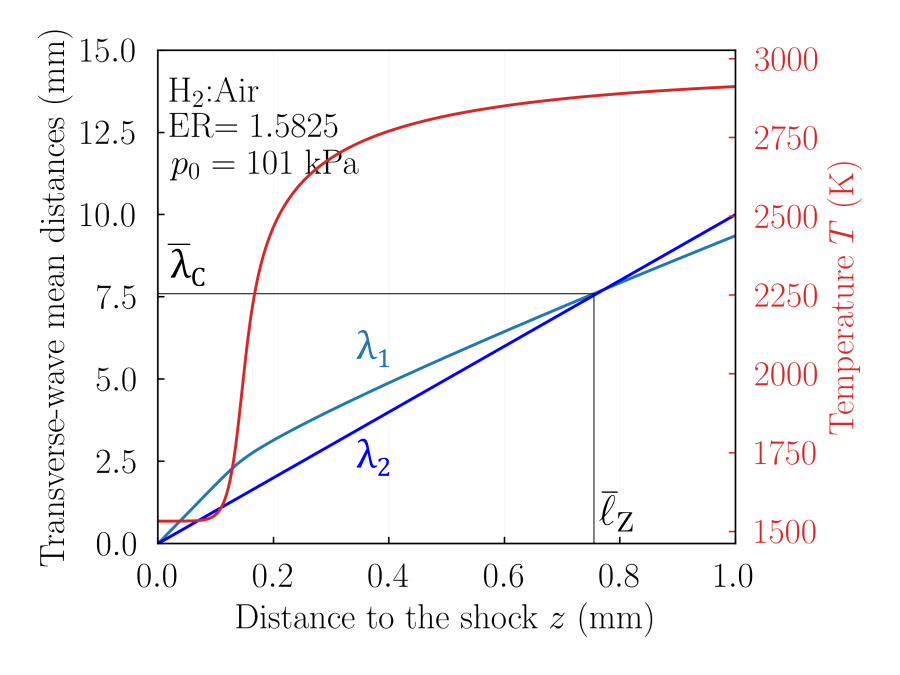}
\end{subfigure}
\begin{subfigure}{0.4\textwidth}
  \centering
  \includegraphics[width=1\linewidth]{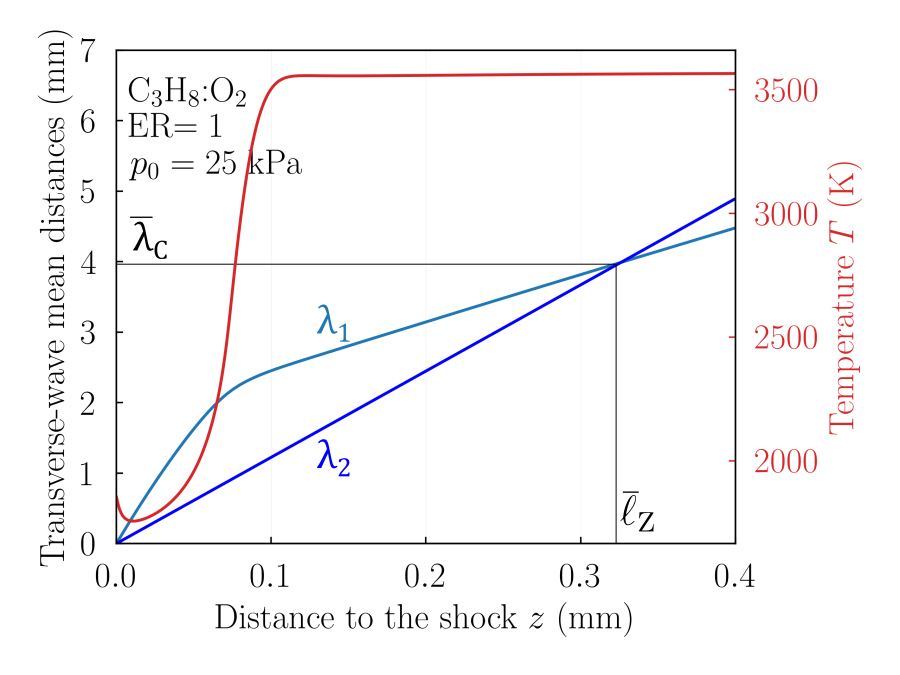}
\end{subfigure}
\begin{subfigure}{0.4\textwidth}
  \centering
  \includegraphics[width=1\linewidth]{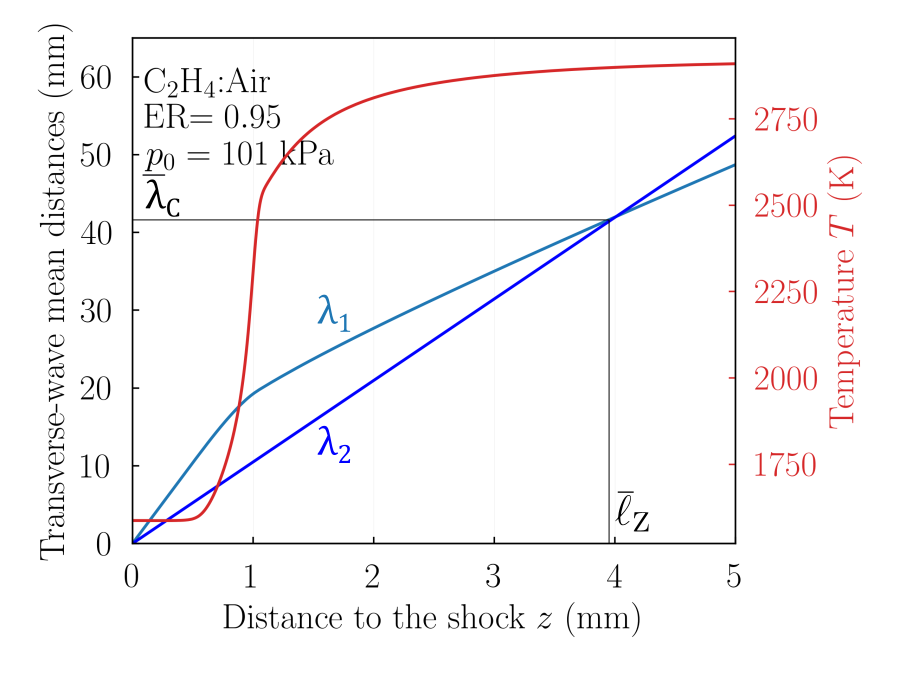}
\end{subfigure}
\caption
{ZND temperature profiles $T$ (red curves, right ordinates) and transverse-wave mean distances ${\lambda}_1$ and ${\lambda}_2$ (blue and green curves, left ordinates). (Konnov's mechanism). The intersection of ${\lambda}_1$ and ${\lambda}_2$ gives the cell mean width $\bar{\lambda}_\text{C}$ and the ZND complete reaction length $\bar{\ell}_\text{Z}$ (\hyperref[sec:Model]{Sects. 3}\hyperref[sec:GraphProba]{ \& 4}).}
\label{fig:Graphs_ZND_profiles}
\end{figure*}

Table \ref{Tab_Check} details data and calculation results for the Konnov's mechanism and the four mixtures \ce{2H2 + O2}, \ce{2H2 + Air}, \ce{C3H8 + 5O2} and \ce{0.95C2H4 + Air}.  The relative differences between experimental and calculated $\bar{\lambda}_\text{C}$ are negligible for all selected cases, confirming, as discussed above, the agreements observed in the figures \ref{fig:Graphs_H2}-a,c,d and \ref{fig:Graphs_CnHm}-a to -c and -f. The mean burnt volume fractions $\bar{\nu}$ (\ref{MeanFracVol}) and the length ratios $\bar{\ell}_\text{ZI}/\delta \bar{\ell}_\text{ZR}$, $\bar{\lambda}_\text{C}/\bar{\ell}_\text{ZI}$ and $\bar{\lambda}_\text{C}/\bar{\ell}_\text{Z}$ differ little from one mixture to another although that they are slightly larger for \ce{C3H8 + 5O2}. The ratios $\bar{\lambda}_\text{C}/\bar{\ell}_{\text{ZI}}$ and $\bar{\lambda}_\text{C}/\bar{\ell}_{\text{Z}}$ have typical large values, \textit{e.g.} \cite{Crane2019}. Figure \ref{fig:Graphs_ZND_profiles} shows the corresponding ZND profiles of the temperatures $T$, and the transverse-wave mean distances ${\lambda}_1$ and ${\lambda}_2$ (\hyperref[sec:Model]{Sect. 2}) whose intersection defines the cell mean width $\bar{\lambda}_\text{C}$ and the ZND complete reaction length $\bar{\ell}_\text{Z}$.

Overall, the model predicts the right values and trends of $\bar{\lambda}_\text{C}$ for cells independent of the confinement
and classified up to irregular provided the kinetic mechanism is physical and suitable for the considered mixture.

\begin{table}[ht!]
\centering
 \caption{Data and calculation results for the Konnov's mechanism \cite{Konnov2007}}
 \label{Tab_Check}
 \renewcommand{\arraystretch}{1.25}
 \begin{tabular}[h!]{c c c c c c}
 \hline
 & & \ce{H2}\,:\,\ce{O2} & \ce{H2}\,:\,Air & \ce{C3H8}\,:\,\ce{O2} & \ce{C2H4}\,:Air \\
\hline
ER &                                            & 1         & 1.5828    & 1         & 0.95 \\
$p_0$ & \footnotesize{kPa}                          & 192       & 101       & 25        & 101 \\
$v_0$ & \footnotesize{m$^3/$kg}                     & 1.056	    &	1.319	&	2.862	&	0.911 \\
$v_\text{N}$ &\footnotesize{m$^3/$kg}               & 0.188	    &	0.211	&	0.268	&	0.146 \\
$v_\text{CJ}$ &\footnotesize{m$^3/$kg}              & 0.575	    &	0.740	&	1.537	&	0.502 \\
$D_\text{CJ}$ &\footnotesize{m/s}              & 2873.2    &	2092.6	&	2296.0	&	1805.8 \\
\hline
$\bar{t}_\text{Z}$ &\footnotesize{$\mu$s}                & 0.079	    &	1.139	&	0.514	&	6.782 \\
$\bar{\ell}_\text{Z}$ & \footnotesize{mm}               & 0.074	    &	0.735	&	0.318	&	3.861 \\
$\bar{\lambda}_\text{C}$&\footnotesize{mm}   & 0.78	&	7.64	&	3.98	&	41.90 \\
$\bar{\lambda}_\text{C}^\text{exp}$&\footnotesize{mm}   & 0.74	&	8.73	&	3.80	&	42.40 \\
\hline
$\bar{v}_\text{Z}$ &\footnotesize{m$^3/$kg}         & 0.337	&	0.414	&	0.756	&	0.283 \\
$\bar{v}_\text{Z}/v_0$ &                            & 0.319	&	0.314	&	0.264	&	0.311 \\
$\bar{\nu}$ & \footnotesize{$\%$vol.}               & 0.657	&	0.687	&	0.782	&	0.682 \\
$\bar{\ell}_\text{Z}(\bar{\nu})$   & \footnotesize{mm}              & 0.073	&	0.749	&	0.312	&	3.810 \\
$\bar{\ell}_\text{ZI}$ &\footnotesize{mm}                 & 0.025	&	0.234	&	0.068	&	1.210 \\
$\delta \bar{\ell}_\text{ZR}$ & \footnotesize{mm}         & 0.048	&	0.515	&	0.244	&	2.600 \\
$\bar{\ell}_\text{ZI}/\delta \bar{\ell}_\text{ZR}$ &            & 0.52  &	0.45	&	0.28	&	0.47 \\
$\bar{t}_\text{ZI}$ & \footnotesize{$\mu$s}               & 0.049	&	0.701	&	0.316	&	4.171 \\
$\delta \bar{t}_\text{ZR}$ & \footnotesize{$\mu$s}        & 0.031	&	0.439	&	0.198	&	2.611 \\
\hline
$\bar{\lambda}_\text{C}/\bar{\ell}_\text{ZI}$& & 31.2	&	32.6	&	58.5	&	34.6 \\
$\bar{\lambda}_\text{C}/\bar{\ell}_\text{Z}$& & 10.7  &	10.2	&	12.8	&	11.0 \\
 \hline
 \end{tabular}
\end{table}

\clearpage

\clearpage
\section{\label{sec:DiscConcl}Discussion and conclusions}

Numerical simulation is acknowledged today as the only tool capable of handling the complex constitutive relationships, such as detailed chemical kinetic mechanisms, required to describe detonation dynamics phenomena, both qualitatively and qualitatively. These phenomena involve essentially compressibility and chemical kinetics, but turbulent diffusion can also play a significant role depending on the mixture. That still requires large computational resources, which precludes this tool from anticipating physically practical situations. Therefore, the mean width $\bar{\lambda}_\text{C}$ of the cells forming the 3D structure of the detonation reaction zones is still used as a detonability indicator useful for device presizing.

Our model for calculating a representative cell width $\bar{\lambda}_\text{C}$ retains the constraint of a detailed kinetic mechanism but uses a statistical approach rather than a detailed description of the complex transverse wave dynamics of the 3D cellular structure. Its object is to readily predict a physical mean width sufficiently accurate for practical purposes. Its physical premise is that the detonation front propagates steadily under conditions such that the cellular and the ZND combustion processes have equivalent average combustion rates (\hyperref[sec:Model]{Sect.3}). Because the ZND model is a representative average only if the number of cells on a 3D front is sufficiently high, the model holds only for multicellular detonations, not marginal ones. Physical multicellular (CJ) detonation regimes show irregular front views. The mean cell widths are independent of the cross-section shape only for sufficiently high initial pressures and large distances from the ignition, \textit{e.g.}, Fig. 7 in \cite{Monnier2022a}. Our model assumes that the front is multicellular in the sense of both limits of long times and large cross-section areas.

Then the stationary ergodicity of the transverse wave dynamics is a fundamental hypothesis in our work (\hyperref[sec:Introduction]{Sects. 1} \hyperref[sec:GraphProba]{\& 4}). Indeed, we know of no experiment that shows statistical distribution of front-view patterns qualitatively different in shape and dimension from one experiment to another if the cell number is sufficiently high. In particular, none of the statistical analyses of the numerous cell recordings made so far could have converged on the same $\bar{\lambda}_\text{C}$ if this were not the case. This statistical description is independent of whether the cellular mechanism is adiabatic.
The representation of an irregular cellular front by a hexagonal patterned tessellation is the limiting rearrangement of the cell edges statistically equivalent to the experimental distribution of the transverse waves at each instant of time.
The constant value $\bar{y}=0.385$ (\ref{MeanFracMass}) of the mean burn mass fraction is the consequence of this ergodic hypothesis and the independence on time of the Descartes-Euler-Poincar\'{e} relation, and its limit. Therefore, we are looking at the interest of an improved representation of the front by a multi-pattern Voronoi tessellation. We investigate the tentative assumption that the transverse wave dynamics is a Poisson random process both stationary and homogeneous, considering that irregular cells, due to their high sensitivity to initial conditions, have center points randomly distributed, statistically constant in number and which move independently of each other. Table \ref{Tab_Poisson_Voronoi} shows the corresponding distribution of the number of edges $E$ of the patterns obtained explicitly by Calka \cite{Calka2003} but recognized implicit in Miles and Maillardet \cite{Miles1982}. We observe the predominance of hexagons, then of pentagons and heptagons, about $75\%$, which supports the representativeness of the hexagon limit used in this work. That also suggests that the mean burnt mass fraction for this multi-pattern tessellation, calculated from a barycentric rule based on the individual mean burnt mass fractions and the distribution of the patterns, would be only slightly different from the value $0.385$ (\ref{MeanFracMass}) for the hexagon alone (\hyperref[sec:GraphProba]{Sect. 4}). We also observe that about $67\%$ of the patterns have at most six edges.
\begin{table}[h!]
 \centering
 \caption{Pattern distribution in $\%$ as a function of the number of edges $E$\\for the homogeneous stationary Poisson process \cite{Miles1982,Calka2003}} \label{Tab_Poisson_Voronoi}
 \begin{tabular}[ht!]{|P{1cm} | P{1cm} P{1cm} P{1cm} P{1cm} P{1cm} P{1cm} P{1cm} P{1cm} | }
 \hline
 $E$ & 3 & 4 & 5 & 6 & 7 & 8 & 9 & 10 \\ [0.5ex] 
 \hline
 $\%$ & 1.1 & 10.7 & 25.9 & 29.5 & 19.9 &  9 &  3 & 0.7 \\
 \hline
 \end{tabular}
\end{table}

Adiabatic heating by shock compression is the ignition mechanism involved in our approach 
to express the physical premise that the ZND and cellular combustion processes should have the same production rate of burnt mass, hence the cellular rate $2/\Bar{t}_\text{C}$ and the relation (\ref{Eq Rate}) \hyperref[sec:Model]{(Sect. 3)}. Together with the adiabatic ZND model (\hyperref[sec:App-ZND]{App. A}), that predicts well $\bar{\lambda}_\text{C}$ for \ce{H2}:\ce{O2}, \ce{H2}:\ce{Air}, \ce{C3H8}:\ce{O2}, \ce{C3H8}:\ce{Air} and \ce{C2H4}:\ce{Air} mixtures but overestimates it for \ce{CH4}:\ce{O2} mixtures (\hyperref[sec:Results]{Sect. 5}). Indeed, the combustion mechanism of the latter mixtures includes turbulent diffusion and yields highly-irregular diamonds. However, all detonations in tubes with sufficiently large transverse dimensions propagate with the CJ velocity, so the premise should be valid regardless of the ignition mechanism. Thus, a possible extension of the present analysis could be to consider the relation (\ref{Eq Rate}) as always valid on the average and to use a non-adiabatic ZND model to investigate whether the model could also predict mean cell widths for highly unstable cellular reaction zones. However, that would neither support nor answer the question of the representativeness of cell mean widths $\bar{\lambda}_\text{C}$ obtained from highly scattered measurements and thus about the sufficiency of a single characteristic length to characterize the geometric properties of longitudinally highly irregular cells.
Indirectly, the adiabatic formulation of the model provides a means of identifying those mixtures whose detonation reaction zone would involve turbulent diffusion.

In conclusion, if the chemical kinetic mechanisms are physical and suitable for the mixtures considered, and based on a simple and rapid post-processing of a ZND result spreadsheet (\hyperref[sec:Model]{Sect. 3}), the model predicts correct values and trends of the mean widths $\bar{\lambda}_\text{C}$ of detonation cells whose longitudinal soot recordings show longitudinally regular to irregular diamonds independent of the confinement.
Possible extensions are an improved representation of the cellular fronts as a multi-pattern Voronoi tessellation governed by a Poisson random process and a coupling with a non-adiabatic ZND reaction zone.

\clearpage
\section*{Acknowledgements} \label{sec:Acknowledgements} 
This work was supported by the Ministry of Higher Education, Research and Innovation, France.

\clearpage  
\appendix
\section{\label{sec:App-ZND}ZND model equations}

We summarize below the basic ZND-model case relevant to this work, namely that for constant-velocity planar detonation in ideal reactive inviscid gases, with detailed chemical kinetic mechanisms. Higgins \cite{Higgins2012_Zhang} and Ng \cite{Ng2012_Zhang} have given comprehensive reviews. The model equations derive from the Euler equations, which express the mass, momentum, and energy balances for the inviscid reactive fluid, simplified by the assumptions that the flow behind the detonation leading shock is one-dimensional planar and steady.

The notation is essentially that introduced in \hyperref[sec:Model]{Section 2} and detailed in the nomenclature (\hyperref[sec:App-Nomen]{App. B}). The origin of the distances $x$ in the laboratory reference frame is an arbitrary initial position $L_0=0$ of the shock and the origin of the times $t$ is the instant $t_0=0$ when the fluid elements enter the ZND reaction zone. Thus, at an instant $t>0$, the position of the shock is $L(t)=Dt$ and the distance relative to this position is $z(x,t)=Dt-x$. Steadiness is the invariance of any variable $f$ with respect to $L(t)$ at a constant relative distance $z$ in the reaction zone. With $f(t,x)$ and $f(t,z)$ the corresponding representations, this gives the constraint
\begin{equation}
\left. \frac{\partial f}{\partial t}\right) _{x}+D\left. \frac{\partial f}{%
\partial x}\right) _{t} =\left. \frac{\partial f}{\partial t}\right)
_{z}=0. \label{Steadns}
\end{equation}
The Euler equations for one-dimensional planar flow
\begin{align}
&\left. \frac{\partial \rho }{\partial t}\right)_{x}+\left. \frac{\partial
\rho u}{\partial x}\right) _{t} =0, \\
&\left. \frac{\partial \rho u}{\partial t}\right)_{x}+\left. \frac{\partial
p+\rho u^{2}}{\partial x}\right)_{t} =0, \\
&\left. \frac{\partial \rho E}{\partial t}\right)_{x}+\left. \frac{\partial
\left( p+\rho E\right) u}{\partial x}\right)_{t} =0,
\end{align}
with $E=e+u^2/2$, $e$ the specific internal energy and $u$ the material speed in the laboratory frame, thus reduce to a system of ordinary differential equations
\begin{equation}
\frac{d\mathbf{f}}{dz}=\mathbf{F}\left(\mathbf{f},D\right) ,\quad \mathbf{f}(z=0)=\mathbf{f}_{\text{H}}(D,p_0,T_0,\mathbf{y}_0), \label{ZND_systG}
\end{equation}
where $d./dz\equiv\left. \partial ./\partial z\right) _{t}$ (since $\left. \partial ./\partial t\right)
_{z}=0$, (\ref{Steadns})) denotes the space derivatives, $\mathbf{f}$ the vector of the dependent variables, namely $p$ the pressure, $\rho$ the specific mass, $U=D-u$ the material speed in the shock frame, $\mathbf{y}$ the vector of the mass fractions $y_{k}$ of the $K$ chemical species ($k=1,K$), $t_\text{m}(z)$ the time relative to the shock for a fluid element to reach the position $z$, and $\mathbf{f}_{\text{H}}$ the values of $\mathbf{f}$ at the shock ($z=0$, $t=0$), which serve as boundary conditions for the integration of the system, given $D$ and the initial pressure $p_0$, temperature $T_0$ and composition $\mathbf{y}_0$.
The system writes
\begin{align}
\frac{dU}{dz} &= \frac{\boldsymbol{\varsigma.}\boldsymbol{\omega}}{1-M_\text{a}^{2}}, \label{dUZ}\\
\frac{dp}{dz} &= -\rho U\frac{\boldsymbol{\varsigma.}\boldsymbol{\omega}}{1-M_\text{a}^{2}}, \label{dpZ}\\
\frac{dv}{dz} &= \frac{1}{\rho U}\frac{\boldsymbol{\varsigma.}\boldsymbol{\omega}}{1-M_\text{a}^{2}}, \label{dvZ}\\
\frac{d\mathbf{y}}{dz} &= \frac{\boldsymbol{\omega}}{U}, \label{dyZ}\\
\frac{dt_\text{m}}{dz} &= \frac{1}{U}, \label{Id_t_z}
\end{align}
where $v=1/\rho$ denotes the specific volume, $M_\text{a}=U/c$ the flow Mach number, $c$ the frozen sound velocity, $\boldsymbol{\omega}=d\mathbf{y}/dt$ the vector of the reaction rates $\omega_k=dy_k/dt$ (with $d./dt=Ud./dz$ the material derivative), $\boldsymbol{\varsigma.}\boldsymbol{\omega}=\dot{\sigma}$ the thermicity, and $\boldsymbol{\varsigma}$ the vector of the thermicity coefficients $\varsigma_k$. The system is formally closed with hydrodynamic constitutive relations, namely the reaction rates $\omega_k(p,v,\mathbf{y})$ and the equation of state $e(p,v,\mathbf{y})$ that defines the frozen sound velocity $c(p,v,\mathbf{y})$ and the thermicity coefficients $\varsigma_k(p,v,\mathbf{y})$ by
\begin{equation}
c^{2}=v^{2}\frac{p+\left. \frac{\partial e}{\partial v}\right) _{p,\mathbf{y}}}{%
\left. \frac{\partial e}{\partial p}\right) _{v,\mathbf{y}}}, \quad \varsigma_{k}=\frac{-v}{c^{2}}\left. \frac{\partial e}{\partial y_{k}}\right) \label{SoundTherm}
_{p,v,y_{j\neq k}}.
\end{equation}
Calculation codes implement thermal equations of state $h(T,p,\mathbf{n})$ and $v(T,p,\mathbf{n})$, with $h=e+pv$ the specific enthalpy and $\mathbf{n}$ the vector of the mole fractions. Chain rule operations transform their derivatives into the hydrodynamic derivatives in (\ref{SoundTherm}) \cite{FickettDavis1979,Ng2012_Zhang,Higgins2012_Zhang}. The joint integration of the identity (\ref{Id_t_z}) with the equations (\ref{dUZ}-\ref{dyZ}) defines the time-position relationships $t_{\text{m}}(z)$ and $z_{\text{m}}(t)$ of a fluid element (\hyperref[sec:Model]{Sect. 2}).

The Euler equations subjected to the steadiness constraint also have first integrals in the form of the Rankine-Hugoniot (RH) relations below, which express the conservation of the mass, momentum and energy surface fluxes from the initial state (subscript $0$) through the shock to any distance $z$ or time $t$ relative to the shock in the reaction zone,
\begin{align}
\rho_0 D&= \rho U, \\
p_0+\rho_0 D^2&= p+\rho U^2, \\
h_0+\frac{1}{2}D^2&= h+\frac{1}{2}U^2.
\end{align}
The boundary condition on the chemical composition at the shock is often the no-dissociation constraint $\mathbf{y}_\text{N}=\mathbf{y}_0$. The RH relations combined with equations of state for $h_\text{N}$ and $h_0$ then determine the boundary value $\mathbf{f}_{\text{H}}(D,p_0,T_0,\mathbf{y}_0)$, and the integration of (\ref{ZND_systG}) then yields the profiles and evolutions of the corresponding dependent variables. The value of $D$ for the Chapman-Jouguet detonation is not arbitrary and is a part of the solution. A preliminary estimate is the equilibrium value $D_\text{CJ}$ obtained by solving the RH relations above supplemented by the chemical equilibrium constraint $\mathbf{A}(T,p,\mathbf{y})=0$ on the chemical affinity $\mathbf{A}$, which defines the composition $\mathbf{y_\text{eq}}(T,p)$ at chemical equilibrium, the equilibrium thermal equations of state $h_\text{eq}(T,p)=h(T,p,\mathbf{y_\text{eq}}(T,p))$ and $v_\text{eq}(T,p)=v(T,p,\mathbf{y_\text{eq}}(T,p))$, and the equilibrium CJ constraint $U=c_\text{eq}(T,p)$, where $c_\text{eq}$ is the equilibrium sound velocity given by
\begin{equation}
c^{2}_\text{eq}=v^{2}\frac{p+\left. \frac{\partial e_\text{eq}}{\partial v}\right) _{p}}{%
\left. \frac{\partial e_\text{eq}}{\partial p}\right) _{v}} \ne (c^{2})_\text{eq}=c^{2}(T,p,\mathbf{y_\text{eq}}(T,p)). \label{Sound_Eq}
\end{equation}
An iterative procedure based on successive integrations of (\ref{ZND_systG}) initiated with $D_\text{CJ}$ then aims at framing to a reasonable degree of accuracy which value of $D$ ensures that the derivatives in the ZND system remain physical, that is, bounded, while approaching the sonic locus, since there $1-M_\text{a}=0$ (or $u+c=D$) and $\boldsymbol{\omega}=\mathbf{0}$ (with $\mathbf{A}(T,p,\mathbf{y})=\mathbf{0}$). That gives the profiles, the evolutions, and the frozen CJ velocity $D_\text{CJ-f}$. The latter is different from the equilibrium velocity $D_\text{CJ}$ because $c_\text{eq} \ne (c)_\text{eq}$ (\ref{Sound_Eq}), but usually not significantly. Higgins \cite{Higgins2012_Zhang} has also discussed the more complex cases of pathological ZND detonations, \textit{e.g.}, $\boldsymbol{\varsigma}\boldsymbol{.\omega}=0$ and $\boldsymbol{\omega}\ne\mathbf{0}$. We implemented the Caltech ZND code with detailed chemical kinetic mechanisms and their thermodynamic database taken from the literature \hyperref[sec:Results]{(Sect. 4)}.

ZND profiles and evolutions easily define theoretical characteristic lengths and times.
Most profiles exhibit essentially two characteristic zones, namely the induction and the reaction, whose thicknesses are used to dimension observable lengths and times of dynamic detonation behaviors, such as average detonation cell widths. There are several definitions of the induction and reaction thicknesses (\hyperref[sec:mod-overview]{Sect. 2}). 
Figure \ref{fig:thermicite} shows an example of ZND temperature and thermicity profiles for the stoichiometric mixture \ce{2H2 + O2} (\hyperref[sec:mod-overview]{Sect. 2}).

\clearpage 
\section{\label{sec:App-Nomen}Nomenclature}


\nomenclature[A,02]{$0$}{Initial state and origin of position and time ($L_0=0$, $t_0=0$)}
\nomenclature[A,03]{1D, 2D, 3D }{One-, two-, and three-dimensional}
\nomenclature[A,06]{CJ}{Chapman-Jouguet}
\nomenclature[A,07]{ZND}{Zel'dovich-von Neuman-Döring}
\nomenclature[A,01]{($\bar{\hspace{0.3cm}}$)}{Mean value}
\nomenclature[A,08]{Z}{ZND process}
\nomenclature[A,09]{C}{Cellular process}
\nomenclature[A,10]{N}{Shock state}
\nomenclature[A,11]{B}{Burnt state}
\nomenclature[A,05]{ER}{Equivalence ratio}
\nomenclature[A,04]{DDB}{Detonation Database}

\nomenclature[B,01]{$\chi$}{Instability parameter}
\nomenclature[B,02]{$\boldsymbol{\varsigma}$}{Vector of the thermicity coefficients \nomunit{}}
\nomenclature[B,03]{$\boldsymbol{\omega}$}{Vector of the reaction rates \nomunit{s$^{-1}$}}
\nomenclature[B,04]{$\dot{\sigma}$}{Thermicity ($\boldsymbol{\varsigma.}\boldsymbol{\omega}$) \nomunit{{s$^{-1}$}}}
\nomenclature[B,05]{$\rho$}{Specific mass ($\rho=1/v$)    \nomunit{kg$\times$m$^{-3}$}}
\nomenclature[B,06]{$\Bar{\nu}$}{Mean burnt volume fraction    \nomunit{}}
\nomenclature[B,07]{$\lambda$}{Transverse-wave distance    \nomunit{m}}
\nomenclature[B,08]{$\bar{\lambda}_\text{C}$}{Cell mean width   \nomunit{m}}
\nomenclature[B,09]{$a$}{Mean cell width-to-length aspect ratio ($\bar{\lambda}_\text{C}/\bar{L}_\text{C} $)   \nomunit{}}
\nomenclature[B,10]{$A_\text{C}$}{Area of the mean cell\nomunit{m$^2$}}
\nomenclature[B,11]{$A_\text{T}$}{Area of the tube\nomunit{m$^2$}}
\nomenclature[B,12]{$c$}{Sound velocity \nomunit{m$\times$s$^{-1}$}}
\nomenclature[B,13]{$D$}{Detonation velocity    \nomunit{m$\times$s$^{-1}$}}
\nomenclature[B,14]{$U$}{Material speed in the ZND shock frame \nomunit{m$\times$s$^{-1}$}}
\nomenclature[B,15]{$d_\text{transv}$}{Transverse dimension of tubes \nomunit{m}}
\nomenclature[B,16]{$L$}{Position of the ZND shock in the laboratory frame   \nomunit{m}}
\nomenclature[B,17]{$\bar{L}_\text{C}$}{Cell mean length    \nomunit{m}}
\nomenclature[B,18]{$\Bar{\ell}_\text{Z}$}{ZND complete reaction length ($\Bar{\ell}_\text{Z}=\bar{U}_\text{Z}\,\Bar{t}_\text{Z}$) \nomunit{m}}
\nomenclature[B,19]{$\Bar{\ell}_\text{ZI}$}{Thickness of the ZND induction layer ($\Bar{\ell}_\text{ZI}=U_\text{N}\,\Bar{t}_\text{ZI}$) \nomunit{m}}
\nomenclature[B,20]{$\delta\Bar{\ell}_\text{ZR}$}{Thickness of the ZND main-reaction layer  ($\delta\Bar{\ell}_\text{ZR}=U_\text{CJ}/\delta\Bar{t}_\text{ZR}$)\nomunit{m}}
\nomenclature[B,21]{$M$}{Mass  \nomunit{kg or kg$\times$m$^{-2}$}}
\nomenclature[B,22]{$M_\text{a}$}{Mach number ($U/c$) \nomunit{}}
\nomenclature[B,23]{$p$}{Pressure     \nomunit{Pa}}
\nomenclature[B,24]{$T$}{Temperature   \nomunit{K}}
\nomenclature[B,25]{$t$}{Time   \nomunit{s}}
\nomenclature[B,26]{$t_\text{m}$}{Material time from the ZND shock  \nomunit{s}}
\nomenclature[B,27]{$\Bar{t}_\text{C}$}{Characteristic time for the length of the mean cell ($\bar{L}_\text{C}/D_\text{CJ}$)   \nomunit{s}}
\nomenclature[B,28]{$\Bar{t}_\text{Z}$}{ZND complete reaction time    \nomunit{s}}
\nomenclature[B,29]{$\Bar{t}_\text{ZI}$}{ZND induction time    \nomunit{s}}
\nomenclature[B,30]{$\delta\Bar{t}_\text{ZR}$}{ZND main-reaction time    \nomunit{s}}
\nomenclature[B,31]{$x$}{Distances in the laboratory frame   \nomunit{m}}
\nomenclature[B,32]{$z$}{Distances measured from the ZND shock    \nomunit{m}}
\nomenclature[B,33]{$\Bar{y}$}{Mean burnt mass fraction    \nomunit{}}
\nomenclature[B,34]{$v$}{Specific volume    \nomunit{m$^3\times$kg$^{-1}$}}

\nomenclature[C,01]{$F$}{Number of faces }
\nomenclature[C,01]{$E$}{Number of edges }
\nomenclature[C,01]{$V$}{Number of vertices }
\nomenclature[C,10]{$d_\text{i}$, $d_\text{x}$}{Inner and outer diameters of a hexagon  \nomunit{m}}

\nomenclature[C,02]{$m_i$}{Non-intersection measure for variable segment length}
\nomenclature[C,02]{$\mu_i$}{Vassallo's non-intersection measure for constant-length segment}
\nomenclature[C,02]{$r$}{Non-dimensional segment length}
\nomenclature[C,02]{$\mu$}{Measure of the total hyper-volume}
\nomenclature[C,02]{$\mu_\text{C}$}{Measure of the non-intersection hyper-volume}

\nomenclature[C,02]{$s$}{Segment length    \nomunit{m}}

\printnomenclature
\clearpage 
\newpage
\bibliographystyle{elsarticle-num}
\bibliography{Bibliography}         

\providecommand{\noopsort}[1]{}\providecommand{\singleletter}[1]{#1}
\begin{thebibliography}{10}
\expandafter\ifx\csname url\endcsname\relax
  \def\url#1{\texttt{#1}}\fi
\expandafter\ifx\csname urlprefix\endcsname\relax\def\urlprefix{URL }\fi
\expandafter\ifx\csname href\endcsname\relax
  \def\href#1#2{#2} \def\path#1{#1}\fi

\bibitem{Denisov1959}
Y.~N. Denisov, Y.~K. Troshin, Pulsating and spinning detonation of gaseous
  mixtures in tubes, Dokl. Akad. Nauk SSSR 125 (1959) 110--113.

\bibitem{Ng2012_Zhang}
H.~D. Ng, Detonation instability, in: F.~Zhang (Ed.), Shock Waves Science and
  Technology Library, Vol. 6: Detonation Dynamics, Springer Berlin Heidelberg,
  2012, pp. 107--212.
\newblock \href {http://dx.doi.org/10.1007/978-3-642-22967-1_3}
  {\path{doi:10.1007/978-3-642-22967-1_3}}.

\bibitem{Clavin2012}
P.~Clavin, F.~A. Williams, Analytical studies of the dynamics of gaseous
  detonations, Philos. Trans. A: Math. Phys. Eng. Sci. 370 (2012) 597--624.
\newblock \href {http://dx.doi.org/10.1098/rsta.2011.0345}
  {\path{doi:10.1098/rsta.2011.0345}}.

\bibitem{Clavin2016}
P.~Clavin, G.~Searby, Combustion waves and fronts in flows: flames, shocks,
  detonations, ablation Fronts and explosion of stars, Cambridge University
  Press, 2016.
\newblock \href {http://dx.doi.org/10.1017/CBO9781316162453}
  {\path{doi:10.1017/CBO9781316162453}}.

\bibitem{Monnier2022a}
V.~Monnier, V.~Rodriguez, P.~Vidal, R.~Zitoun, An analysis of three-dimensional
  patterns of experimental detonation cells, Combust. Flame 245 (2022) 112310.
\newblock \href {http://dx.doi.org/10.1016/j.combustflame.2022.112310}
  {\path{doi:10.1016/j.combustflame.2022.112310}}.

\bibitem{Crane2023}
J.~Crane, J.~T. Lipkowicz, X.~Shi, I.~Wlokas, A.~M. Kempf, H.~Wang,
  Three-dimensional detonation structure and its response to confinement, Proc.
  Combust. Inst. 39~(3) (2023) 2915--2923.
\newblock \href {http://dx.doi.org/10.1016/j.proci.2022.10.019}
  {\path{doi:10.1016/j.proci.2022.10.019}}.

\bibitem{Soloukhin1969}
R.~I. Soloukhin, Nonstationary phenomena in gaseous detonation, Symp. (Int.)
  Combust. 12~(1) (1969) 799--807.
\newblock \href {http://dx.doi.org/10.1016/S0082-0784(69)80461-3}
  {\path{doi:10.1016/S0082-0784(69)80461-3}}.

\bibitem{Subbotin1975}
V.~A. Subbotin, Two kinds of transverse wave structures in multifront
  detonation, Combustion, Explosion and Shock Waves 11 (1975) 83--88.
\newblock \href {http://dx.doi.org/10.1007/BF00742862}
  {\path{doi:10.1007/BF00742862}}.

\bibitem{Radulescu2005}
M.~I. Radulescu, G.~J. Sharpe, J.~H.~S. Lee, C.~B. Kiyanda, A.~J. Higgins,
  R.~K. Hanson, The ignition mechanism in irregular structure gaseous
  detonations, Proc. Combust. Inst. 30 (2005) 1859--1867.
\newblock \href {http://dx.doi.org/10.1016/j.proci.2004.08.047}
  {\path{doi:10.1016/j.proci.2004.08.047}}.

\bibitem{Meagher2023}
P.~A. Meagher, X.~Shi, J.~P. Santos, N.~K. Muraleedharan, J.~Crane, A.~Y.
  Poludnenko, H.~Wang, X.~Zhao, Isolating gasdynamic and chemical effects on
  the detonation cellular structure: A combined experimental and computational
  study, Proc. Combust. Inst. 39~(3) (2023) 2865--2873.
\newblock \href {http://dx.doi.org/10.1016/j.proci.2022.08.001}
  {\path{doi:10.1016/j.proci.2022.08.001}}.

\bibitem{Monnier2022b}
V.~Monnier, V.~Rodriguez, P.~Vidal, R.~Zitoun, Experimental analysis of
  cellular detonation: a discussion on regularity and three-dimensional
  patterns, Proc. $28^{\text{th}}$ Int. Coll. Dynamics Explosions Reactive
  Systems (2022), Paper 57.

\bibitem{StrehlowEngel1969}
R.~A. Strehlow, C.~D. Engel, {Transverse waves in detonations: II. Structure
  and spacing in \ce{H_2-O_2}, \ce{C_2H_2-O_2}, \ce{C_2H_4-O_2} and
  \ce{CH_4-O_2} systems}, AIAA J. 7 (1969) 492--496.
\newblock \href {http://dx.doi.org/10.2514/3.5134} {\path{doi:10.2514/3.5134}}.

\bibitem{Libouton1981}
J.~C. Libouton, A.~Jacques, P.~J. Van~Tiggelen, Cinétique, structure et
  entretien des ondes de détonation, Act. Coll. Int.
  Berthelot-Vieille-Mallard-Le Chatelier (1981) 437--442.

\bibitem{Sharpe2008}
G.~J. Sharpe, J.~J. Quirk, Nonlinear cellular dynamics of the idealized
  detonation model: Regular cells, Combust. Theor. Model. 12 (2008) 1--21.
\newblock \href {http://dx.doi.org/10.1080/13647830701335749}
  {\path{doi:10.1080/13647830701335749}}.

\bibitem{Desbordes2012_Zhang}
D.~Desbordes, H.-N. Presles, Multi-scaled cellular detonation, in: F.~Zhang
  (Ed.), Shock Waves Science and Technology Library, Vol. 6: Detonation
  Dynamics, Springer Berlin Heidelberg, 2012, pp. 281--338.
\newblock \href {http://dx.doi.org/10.1007/978-3-642-22967-1_5}
  {\path{doi:10.1007/978-3-642-22967-1_5}}.

\bibitem{Zhao2016}
H.~Zhao, J.~H. Lee, J.~Lee, Y.~Zhang, Quantitative comparison of cellular
  patterns of stable and unstable mixtures, Shock Waves 26~(10) (2016)
  621--633.

\bibitem{Higgins2012_Zhang}
A.~Higgins, Steady one-dimensional detonations, in: F.~Zhang (Ed.), Shock Waves
  Science and Technology Library, Vol. 6: Detonation Dynamics, Springer Berlin
  Heidelberg, 2012, pp. 33--105.
\newblock \href {http://dx.doi.org/10.1007/978-3-642-22967-1_2}
  {\path{doi:10.1007/978-3-642-22967-1_2}}.

\bibitem{Short2003}
M.~Short, G.~J. Sharpe, Pulsating instability of detonations with a two-step
  chain-branching reaction model: theory and numerics, Combust. Theor. Model. 7
  (2003) 401--416.
\newblock \href {http://dx.doi.org/10.1088/1364-7830/7/2/311}
  {\path{doi:10.1088/1364-7830/7/2/311}}.

\bibitem{Radulescu2003}
M.~I. Radulescu, The propagation and failure mechanism of gaseous detonations:
  experiments in porous-waled tubes, PhD. thesis, McGill University Libraries,
  2003.

\bibitem{Radulescu2013}
M.~I. Radulescu, G.~J. Sharpe, D.~Bradley, A universal parameter quantifying
  explosion hazards, detonability and hot spot formation: the $\chi$ number,
  in: Proc. 7$^\text{th}$ Int. Seminar on Fire and Explosion Hazards, Research
  Publishing, 2013, pp. 617--626.
\newblock \href {http://dx.doi.org/10.3850/978-981-07-5936-0_10-01}
  {\path{doi:10.3850/978-981-07-5936-0_10-01}}.

\bibitem{Stamps1991}
D.~W. Stamps, S.~R. Tieszen, The influence of initial pressure and temperature
  on hydrogen-air-diluent detonations, Combust. Flame 83 (1991) 353--364.
\newblock \href {http://dx.doi.org/10.1016/0010-2180(91)90082-M}
  {\path{doi:10.1016/0010-2180(91)90082-M}}.

\bibitem{Auffret1999}
Y.~Auffret, D.~Desbordes, H.-N. Presles, Detonation structure of
  \ce{C2H4-O2-Ar} mixtures at elevated initial temperature, Shock Waves 9
  (1999) 107–111.
\newblock \href {http://dx.doi.org/10.1007/s001930050145}
  {\path{doi:10.1007/s001930050145}}.

\bibitem{Auffret2001}
Y.~Auffret, D.~Desbordes, H.-N. Presles, Detonation structure and detonability
  of \ce{C2H2-O2} mixtures at elevated initial temperature, Shock Waves 11
  (2001) 89–96.
\newblock \href {http://dx.doi.org/10.1007/PL00004069}
  {\path{doi:10.1007/PL00004069}}.

\bibitem{Manzhalei1974}
V.~I. Manzhalei, V.~V. Mitrofanov, V.~A. Subbotin, Measurement of
  inhomogeneities of a detonation front in gas mixtures at elevated pressures,
  Combustion, Explosion and Shock Waves 10 (1974) 89--95.
\newblock \href {http://dx.doi.org/10.1007/BF01463793}
  {\path{doi:10.1007/BF01463793}}.

\bibitem{Bull1982}
D.~Bull, J.~Elsworth, P.~Shuff, E.~Metcalfe, Detonation cell structures in
  fuel/air mixtures, Combustion and Flame 45 (1982) 7--22.
\newblock \href {http://dx.doi.org/10.1016/0010-2180(82)90028-1}
  {\path{doi:10.1016/0010-2180(82)90028-1}}.

\bibitem{Moen1984}
I.~Moen, J.~Funk, S.~Ward, G.~Rude, P.~Thibault, Detonation length scales for
  fuel-air explosives, Prog. Astronaut. Aeronaut. 94 (1984) 55--79.
\newblock \href {http://dx.doi.org/10.2514/5.9781600865695.0055.0079}
  {\path{doi:10.2514/5.9781600865695.0055.0079}}.

\bibitem{Tieszen1986}
S.~Tieszen, M.~Sherman, W.~Benedick, J.~Shepherd, Detonation cell size
  measurements in hydrogen-air-stream mixtures, in: J.-C. Leyer, R.~Soloukhin,
  J.~Bowen (Eds.), Dynamics of Explosions, American Institute of Aeronautics
  and Astronautics, 1986, pp. 205--219.
\newblock \href {http://dx.doi.org/10.2514/5.9781600865800.0205.0219}
  {\path{doi:10.2514/5.9781600865800.0205.0219}}.

\bibitem{Aminallah1993}
M.~Aminallah, J.~Brossard, A.~A. A.~Vassiliev, Cylindrical detonations in
  methane-oxygen-nitrogen mixtures, Prog. Astronaut. Aeronaut. 153 (1984)
  203--228.
\newblock \href {http://dx.doi.org/10.2514/5.9781600866265.0203.0228}
  {\path{doi:10.2514/5.9781600866265.0203.0228}}.

\bibitem{Pintgen2003}
F.~P. Pintgen, J.~E. Shepherd, Simultaneous soot foil and plif imaging of
  propagating detonations, Proc. $19^{\text{th}}$ Int. Coll. Dynamics
  Explosions Reactive Systems (2003), Paper 119.

\bibitem{Pintgen2003b}
F.~P. Pintgen, C.~A. Eckett, J.~M. Austin, J.~E. Shepherd, Direct observations
  of reaction zone structure in propagating detonations, Combust. Flame 133
  (2003) 211--229.
\newblock \href {http://dx.doi.org/10.1016/S0010-2180(02)00458-3}
  {\path{doi:10.1016/S0010-2180(02)00458-3}}.

\bibitem{Austin2005}
J.~M. Austin, F.~P. Pintgen, J.~E. Shepherd, Reaction zones in highly unstable
  detonations, Proc. Combust. Inst. 30 (2005) 1849--1857.
\newblock \href {http://dx.doi.org/10.1016/j.proci.2004.08.157}
  {\path{doi:10.1016/j.proci.2004.08.157}}.

\bibitem{Frederick2022}
M.~D. Frederick, R.~M. Gejji, J.~E. Shepherd, C.~D. Slabaugh, Time-resolved
  imaging of the cellular structure of methane and natural gas detonations,
  Shock Waves 32 (2022) 337–351.
\newblock \href {http://dx.doi.org/10.1007/s00193-022-01080-8}
  {\path{doi:10.1007/s00193-022-01080-8}}.

\bibitem{Vassiliev2012_Zhang}
A.~A. Vasil'ev, Dynamics parameters of detonation, in: F.~Zhang (Ed.), Shock
  Waves Science and Technology Library, Vol. 6: Detonation Dynamics, Springer
  Berlin Heidelberg, 2012, pp. 213--279.
\newblock \href {http://dx.doi.org/10.1007/978-3-642-22967-1_4}
  {\path{doi:10.1007/978-3-642-22967-1_4}}.

\bibitem{Crane2022}
J.~Crane, J.~T. Lipkowicz, X.~Shi, I.~Wlokas, A.~M. Kempf, H.~Wang,
  Three-dimensional detonation structure and its response to confinement, Proc.
  Combustion Institute\href {http://dx.doi.org/10.1016/j.proci.2022.10.019}
  {\path{doi:10.1016/j.proci.2022.10.019}}.

\bibitem{Crane2019}
J.~Crane, X.~Shi, A.~V. Singh, Y.~Tao, H.~Wang, Isolating the effect of
  induction length on detonation structure: Hydrogen–oxygen detonation
  promoted by ozone, Combustion and Flame 200 (2019) 44--52.
\newblock \href {http://dx.doi.org/10.1016/j.combustflame.2018.11.008}
  {\path{doi:10.1016/j.combustflame.2018.11.008}}.

\bibitem{JoulinVidal1998}
G.~Joulin, P.~Vidal, An introduction to the instability of flames, shocks, and
  detonations, in: C.~Godrèche, P.~E. Manneville (Eds.), Hydrodynamics and
  Nonlinear Instabilities, Collection Alea-Saclay: Monographs and Texts in
  Statistical Physics, Cambridge University Press, 1998, p. 493–674.
\newblock \href {http://dx.doi.org/10.1017/CBO9780511524608.007}
  {\path{doi:10.1017/CBO9780511524608.007}}.

\bibitem{Clavin2002}
P.~Clavin, B.~Denet, Diamond patterns in the cellular front of an overdriven
  detonation, Phys. Rev. Lett. 88 (2002) 044502.
\newblock \href {http://dx.doi.org/10.1103/PhysRevLett.88.044502}
  {\path{doi:10.1103/PhysRevLett.88.044502}}.

\bibitem{Barthel1971}
H.~O. Barthel, Reaction zone‐shock front coupling in detonations, Phys.
  Fluids 15~(1) (1972) 43--50.
\newblock \href {http://dx.doi.org/10.1063/1.1693753}
  {\path{doi:10.1063/1.1693753}}.

\bibitem{Barthel1974}
H.~O. Barthel, Predicted spacings in hydrogen‐oxygen‐argon detonations,
  Phys. Fluids 17~(8) (1974) 1547--1553.
\newblock \href {http://dx.doi.org/10.1063/1.1694932}
  {\path{doi:10.1063/1.1694932}}.

\bibitem{Strehlow1967}
R.~Strehlow, R.~Liaugminas, R.~Watson, J.~Eyman, Transverse wave structure in
  detonations, Proc. Combust. Inst. 11~(1) (1967) 683--692.
\newblock \href
  {http://dx.doi.org/https://doi.org/10.1016/S0082-0784(67)80194-2}
  {\path{doi:https://doi.org/10.1016/S0082-0784(67)80194-2}}.

\bibitem{Strehlowetal1969}
R.~A. Strehlow, R.~E. Maurer, R.~S., {Transverse waves in detonations: I.
  Spacing in the Hydrogen-Oxygen System}, AIAA J. 7 (1969) 323--328.
\newblock \href {http://dx.doi.org/10.2514/3.5093} {\path{doi:10.2514/3.5093}}.

\bibitem{Crane2021}
J.~Crane, X.~Shi, J.~T. Lipkowicz, A.~M. Kempf, H.~Wang, Geometric modeling and
  analysis of detonation cellular stability, Proc. Combust. Inst. 38~(3) (2021)
  3585--3593.
\newblock \href {http://dx.doi.org/10.1016/j.proci.2020.06.278}
  {\path{doi:10.1016/j.proci.2020.06.278}}.

\bibitem{Cheevers2022}
K.~Cheevers, M.~Radulescu, Ignition behind decaying shock waves : Detonation
  cells (2022).
\newblock \href {http://arxiv.org/abs/2211.05216} {\path{arXiv:2211.05216}}.

\bibitem{Volin1960}
V.~A. Volin, Y.~K. Troshin, G.~I. Filatov, K.~I. Shchelkin, On the reactional
  kinetic nature of non-uniformities in the shock front and their role in the
  process of gaseous detonation propagation, J. Appl. Mech. and Tech. Phys. 2
  (1960) 78--89.

\bibitem{Shepherd1986}
J.~Shepherd, Chemical kinetics of hydrogen-air-diluent detonations, in: J.-C.
  Leyer, R.~Soloukhin, J.~Bowen (Eds.), Dynamics of Explosions, Vol. 106,
  Progress in Astronautics and Aeronautics Series, 1986, pp. 263--293.
\newblock \href {http://dx.doi.org/10.2514/5.9781600865800.0263.0293}
  {\path{doi:10.2514/5.9781600865800.0263.0293}}.

\bibitem{Gavrikov2000}
A.~I. Gavrikov, A.~A. Efimenko, S.~B. Dorofeev, A model for detonation cell
  size prediction from chemical kinetics, Combust. Flame 120 (2000) 19--33.
\newblock \href {http://dx.doi.org/10.1016/S0010-2180(99)00076-0}
  {\path{doi:10.1016/S0010-2180(99)00076-0}}.

\bibitem{VanTiggelen1989}
P.~J. Van~Tiggelen, J.~C. Libouton, Evolution des variables chimiques et
  physiques à l'intérieur d'une maille de détonation, Ann. Phys. 14 (1989)
  649--660.
\newblock \href {http://dx.doi.org/10.1051/anphys:01989001406064900}
  {\path{doi:10.1051/anphys:01989001406064900}}.

\bibitem{Vassallo2021}
S.~F. Vassallo, Buffon's coin and needle problems for the snub hexagonal
  tiling, Adv. Math.: Sci. J. 10 (2021) 2223--2233.
\newblock \href {http://dx.doi.org/0.37418/amsj.10.4.36}
  {\path{doi:0.37418/amsj.10.4.36}}.

\bibitem{Takai1975}
R.~Takai, K.~Yoneda, T.~Hikita, Study of detonation wave structure, Proc.
  Combust. Inst. 15 (1975) 69--78.
\newblock \href {http://dx.doi.org/10.1016/S0082-0784(75)80285-2}
  {\path{doi:10.1016/S0082-0784(75)80285-2}}.

\bibitem{DATABASE}
M.~Kaneshige, J.~Shepherd,
  \href{https://shepherd.caltech.edu/detn_db/html/db.html}{Detonation
  database}, {California Institute of Technology (2002), Last accessed March
  2023}.
\newline\urlprefix\url{https://shepherd.caltech.edu/detn_db/html/db.html}

\bibitem{Konnov2007}
F.~H.~V. Coppens, J.~{De Ruyck}, A.~A. Konnov, {The effects of composition on
  burning velocity and nitric oxide formation in laminar premixed flames of
  \ce{CH4 + H2 + O2 + N2}}, Combust. Flame 149 (2007) 409--417.
\newblock \href {http://dx.doi.org/10.1016/j.combustflame.2007.02.004}
  {\path{doi:10.1016/j.combustflame.2007.02.004}}.

\bibitem{SanDiego}
\href{https://web.eng.ucsd.edu/mae/groups/combustion/mechanism.html}{{Chemical-Kinetic
  Mechanisms for Combustion Applications, San Diego Mechanism web page,
  Mechanical and Aerospace Engineering (Combustion Research), University of
  California at San Diego (2016), Last accessed March 2023}}.
\newline\urlprefix\url{https://web.eng.ucsd.edu/mae/groups/combustion/mechanism.html}

\bibitem{Smith2016}
G.~P. Smith, Y.~Tao, H.~Wang,
  \href{https://web.stanford.edu/group/haiwanglab/FFCM1/pages/download.html}{{Foundational
  Fuel Chemistry Model Version 1.0 (FFCM-1) (2016), Last accessed March 2023}}.
\newline\urlprefix\url{https://web.stanford.edu/group/haiwanglab/FFCM1/pages/download.html}

\bibitem{Konnov2019}
A.~A. Konnov, Yet another kinetic mechanism for hydrogen combustion, Combust.
  Flame 203 (2019) 14--22.
\newblock \href {http://dx.doi.org/10.1016/j.combustflame.2019.01.032}
  {\path{doi:10.1016/j.combustflame.2019.01.032}}.

\bibitem{Calka2003}
P.~Calka, {An explicit expression of the distribution of the number of sides of
  the typical Poisson-Voronoi cell}, J. Appl. Probab. 35 (2003) 863--870.
\newblock \href {http://dx.doi.org/10.1239/aap/1067436323}
  {\path{doi:10.1239/aap/1067436323}}.

\bibitem{Miles1982}
R.~E. Miles, R.~J. Maillardet, {The basic structures of Voronoi and generalized
  Voronoi polygons}, J. Appl. Probab. 19 (1982) 97--111.
\newblock \href {http://dx.doi.org/10.2307/3213553}
  {\path{doi:10.2307/3213553}}.

\bibitem{FickettDavis1979}
W.~Fickett, W.~C. Davis, Detonation, University of California Press, Ltd.,
  London, 1979.
\newblock \href {http://dx.doi.org/10.1002/prep.19810060307}
  {\path{doi:10.1002/prep.19810060307}}.

\end{thebibliography}

\end{document}